\documentclass[reprint,showpacs,prd,preprintnumbers,amssymb,nofootinbib,superscriptaddress,aps,onecolumn]{revtex4}
\usepackage{graphicx,floatflt,amssymb,rotate,epsfig}
\usepackage[utf8]{inputenc}
\usepackage[inline]{enumitem}
\usepackage{mathrsfs}
\usepackage{amssymb}
\usepackage{amsmath}
\usepackage{color}
\usepackage{graphicx}
\usepackage{subfig}
\usepackage{multirow}
\usepackage{mathtools}
\usepackage{float}
\usepackage{pstricks}
\usepackage[toc,page]{appendix}
\usepackage{pifont}
\usepackage{braket}
\usepackage{bbold}
\usepackage{hyperref}
\usepackage{relsize}
\def\BaBar{{\mbox{\slshape B\kern-0.1em{\smaller A}\kern-0.1em B\kern-0.1em{\smaller A\kern-0.2em R}}}}

\newcommand{\ba}{\begin{array}}
	\newcommand{\ea}{\end{array}}
\def\beq{\begin{equation}}
\def\eeq{\end{equation}}
\def\bea{\begin{eqnarray}}
\def\eea{\end{eqnarray}}

\def\roughly#1{\mathrel{\raise.3ex\hbox
		{$#1$\kern-.75em\lower1ex\hbox{$\sim$}}}}
\def\lsim{\roughly<}
\def\gsim{\roughly>}
\def\sla#1{\raise.15ex\hbox{$/$}\kern-.57em #1}% Feynman slash

\def\bd{B_d^0}

\def\order{\lower 1.8ex \hbox{\LARGE\~{}}}
\def\btopilnu{B \to \pi\ell\nu}
\def\btopiplnu{B^0 \to \pi^+ \ell\nu}
\def\btopi0lnu{B^+ \to \pi^0 \ell\nu}
\def\btorhoplnu{B^0 \to \rho^+ \ell\nu}
\def\btorho0nu{B^+ \to \rho^0 \ell\nu}
\def\btorholnu{B \to \rho \ell\nu}

\newcommand*{\rom}[1]{\expandafter\@slowromancap\romannumeral #1@}

\def\bd0tau{B\to D \tau\nu_{\tau}}

\def\be {\begin{equation}}
\def\ee {\end{equation}}

\usepackage[normalem]{ulem}

\definecolor{darkgreen}{cmyk}{1,0,1,0.4}

\definecolor{pink}{cmyk}{0.4,1,0.3,0}

\def\com2#1{\textcolor{red}{\it{#1}}}
\begin{document}
	%opening
	\title{A closer look at the extraction of $|V_{ub}|$ from $\btopilnu$}
	
	\author{Aritra Biswas}
	\email{iluvnpur@gmail.com}
	\affiliation{Indian Institute of Technology, North Guwahati, Guwahati 781039, Assam, India }
	
	\author{Soumitra Nandi}
	\email{soumitra.nandi@iitg.ernet.in}
	\affiliation{Indian Institute of Technology, North Guwahati, Guwahati 781039, Assam, India }
	
	\author{Sunando Kumar Patra}
	\email{sunando.patra@gmail.com}
	\affiliation{Department of Physics, Bangabasi Evening College, 19 Rajkumar Chakraborty Sarani, Kolkata, 700009, West Bengal, India }
	
	\author{Ipsita Ray}
	\email{ipsitaray02@gmail.com}
	\affiliation{Indian Institute of Technology, North Guwahati, Guwahati 781039, Assam, India }

\begin{abstract}  
	To extract the Cabibbo-Kobayashi-Maskawa (CKM) matrix element $|V_{ub}|$, we have re-analyzed all the available inputs (data and theory) on the $\btopilnu$ decays including the newly available inputs on the form-factors from light cone sum rule (LCSR) approach. We have reproduced and compared the results with the procedure taken up by the Heavy Flavor Averaging Group (HFLAV), while commenting on the effect of outliers on the fits. After removing the outliers and creating a comparable group of data-sets, we mention a few scenarios in the extraction of $|V_{ub}|$. In all those scenarios, the extracted values of $|V_{ub}|$ are higher than that obtained by HFLAV. Our best results for $|V_{ub}|^{exc.}$ are $(3.94 \pm 0.14)\times 10^{-3}$ and $(3.93_{-0.15}^{+0.14})\times 10^{-3}$ in frequentist and Bayesian approaches, respectively, which are consistent with that extracted from inclusive decays $|V_{ub}|^{inc}$ within $1~\sigma$ confidence interval.   
\end{abstract}   
	
	\maketitle
%%%%%%%%%%%%%%%%%%%%%%%%%%%%%%%%%%%%%%%%
\section{Introduction}
%%%%%%%%%%%%%%%%%%%%%%%%%%%%%%%%%%%%%%%%
The tree level semileptonic $b\to u\ell\nu_{\ell}$ ($\ell = e, \mu$) decays are useful probes for extracting the CKM element $|V_{ub}|$. In this regard, both exclusive decays ($\btopilnu$), and inclusive decays ($B \to X_u \ell\nu_{\ell}$) play important roles. At present the extracted values are given by \cite{pdg}
\begin{equation}
|V_{ub}|^{exc} = (3.70 \pm 0.16) \times 10^{-3},\ \ \ \ \text{and}\ \ |V_{ub}|^{inc} = (4.25 \pm 0.12^{+0.15}_{-0.14}) \times 10^{-3},
\end{equation} 
which are in mutual disagreement (by $\gsim 2.2~\sigma$). Here, we would like to mention that unlike the inclusive determination of $|V_{cb}|$ from $B\to X_c\ell\nu_{\ell}$ the inclusive determination of $|V_{ub}|$ is not very clean. The main issue is the large background from the $b\to c\ell\nu$ decays and experimental cuts are necessary to distinguish $b\to u$ transitions from $b\to c$ which restrict the phase-space region where the decay-rate is measured. This complicates the theoretical interpretation of the measurement since it forces us to a corner of the phase-space where usual operator product expansion (OPE) breaks down. Instead of the heavy quark expansion (HQE) parameters, the dependence on the QCD shape functions become more important, which, as a result, enhances the sensitivity to the non-perturbative aspects of the decay. There are different approaches to model the shape function, which renders the extracted values of $|V_{ub}|$ model dependent. For a brief review see \cite{pdg}. $|V_{ub}|$ values, obtained from these approaches, differ from each other. In a very recent analysis of the inclusive spectra with hadronic-tagging, Belle has extracted the values of $|V_{ub}|$ by four different methods, like BLNP \cite{BLNP2005}, DGE \cite{DGE2005}, GGOU \cite{GGOU2007}, and ADFR \cite{ADFR}. By taking an arithmetic average of these four different values from the four different methods, they obtain
\begin{equation}
|V_{ub}|^{inc} = (4.10 \pm 0.09\pm 0.22 \pm 0.15)\times 10^{-3}\,.
\end{equation}
This is the most precise measurement till date.   

The extractions of $|V_{ub}|$ from $\btopilnu$ is also not very clean. The methodology adopted by the Heavy Flavor Averaging Group (HFLAV) is not free from caveats. HFLAV carries out a two-stage process for the extraction of $|V_{ub}|^{exc.}$. In the first stage, they obtain an average squared four-momentum transfer ($q^2$) spectrum from a binned maximum-likelihood fit to determine the average partial branching fraction in each $q^2$ interval. For this fit, they use the available data on the differential $\btopilnu$ decay rates from BaBar(11)~\cite{delAmoSanchez:2010af}, Belle(11)~\cite{Ha:2010rf}, BaBar(12)~\cite{Lees:2012vv}, and Belle(13)~\cite{Sibidanov:2013rkk}. As presented in their review \cite{Amhis:2019ckw}, the quality of this fit is not good, and the $p$-value is around $6$\%. 
In the second fit, this average $q^2$ spectrum along with the lattice and LCSR (at $q^2=0$) inputs had been used to extract $|V_{ub}|$; see ref. \cite{Amhis:2019ckw}. This is a reasonably good fit with $p$-value $\sim 47\%$. The available data and the average $q^2$-spectrum as obtained by HFLAV and the second fit results have been summarized in section 6.3.1 of \cite{Amhis:2019ckw}. After repeating a similar fit mentioned above to obtain the average $q^2$ spectrum, we have arrived at an even worse quality of fit with a $p$ value $<1\%$, though we have used the same experimental information as HFLAV. In any case, a frequentist fit of probability $\lsim 5\%$ is usually considered to be of negligible significance and any further fit (in the second stage), using the outcome of this very low-significance fit is bound to churn out biased predictions for $|V_{ub}|$. It thus becomes essential to reconsider other possible ways of analyzing the available data and pin-point the source of tension in the fits. We discuss the details in the next section.  Subsequently, we have utilized the newly available inputs on the form-factors from LCSR for non-zero values of $q^2$ in our fits.

\subsection{Motivation and a few observations }
%%%%%%%%%%%%%%%%%%%%%%%%%%%%%%%%%%%%%%%%
\subsubsection{Theoretical Background}\label{subsec:theory}

The differential decay width w.r.t. $q^2$ for a pseudoscalar to pseudoscalar semileptonic decay is a function of $f_{+,0}(q^2)$. In particular, for $\bar{B^0}\to\pi^+$ semileptonic transitions, we have\footnote{The corresponding charged $B$ will decay to a neutral pion and hence will be scaled by a factor of $1/2$ at the decay width level since $\pi^0=\frac{u\bar{u}-d\bar{d}}{\sqrt{2}}$}:
\begin{align}
\frac{d\Gamma}{dq^2}\left(\bar{B^0}\to\pi^+l^-\bar{\nu}_l\right)
=  &\frac{G_F^2|V_{ub}|^2}{24\pi^3m_{B^0}^2q^4}\left(q^2-m_l^2\right)^2\left|p_{\pi}(m_{B^0},m_{\pi^+},q^2)\right|\times \nonumber\\ 
&\left[\left(1+\frac{m_l^2}{2q^2}\right)m_{B^0}^2 \left|p_{\pi}(m_{B^0},m_{\pi^+},q^2)\right|^2\left|f_+\left(q^2\right)\right|^2 +
\frac{3m_l^2}{8q^2}\left(m_{B^0}^{2}-m_{\pi^+}^{2}\right)^2\left|f_0\left(q^2\right)\right|^2\right].
\end{align}
where $\left|p_\pi(m_B,m_\pi,q^2)\right|=\sqrt{\lambda(m_B,m_\pi,q^2)}/2 m_B$ with $\lambda(m_B,m_\pi,q^2)=((m_B-m_\pi)^2 - q^2)((m_B + m_\pi)^2 - q^2)$. Therefore, to extract $|V_{ub}|$, we need information on the form-factors at different values of $q^2$. There are non-perturbative techniques like lattice-QCD and LCSR which can provide the values of the form-factors at high and low values of $q^2$, respectively. At present the lattice estimates are available on $f_{+/0}(q^2)$ at zero and non-zero recoils \cite{Flynn:2015mha,Lattice:2015tia,Gelzer:2019zwx}. There is also a recent update on the values of these form-factors at $q^2=0$ and at values other than $q^2=0$ \cite{Gubernari:2018wyi}. So far, inputs from lattice-QCD, experimentally measured values of the decay rates, and LCSR input at $q^2=0$ have been utilized for extracting $|V_{ub}|$. Here, the major sources of uncertainties in the extractions of $|V_{ub}|$ are the form factors. 

To get the shape of the decay rate distribution, one needs to know the shape of the corresponding form-factors in the whole $q^2$ region. Therefore, it is crucial to have a parametrization of $f_{+/0}(q^2)$ that satisfies real analyticity in the complex $q^2$ plane. For the form-factor parametrization, we have followed two different approaches which are known as Bourrely-Caprini-Lellouch (BCL)~\cite{BCL} and Bharucha-Straub-Zwicky (BSZ)~\cite{Straub:2015ica} parametrization and compared their results. 

According to BCL, $f_+$ and $f_0$ are as follows:
\begin{equation}\label{eq:bclexpfp}
f_+(z) = \frac{1}{1 - q^2/m_{B^*}^2} \sum_{n=0}^{N_z-1} b_n^+ \, [z^n-(-1)^{n-N_z}\frac{n}{N_z} z^{N_z}]\,,
\end{equation}
\begin{equation}\label{eq:bclexpf0}
f_0(z) = \sum_{n=0}^{N_z-1} b_n^0 z^n\,.
\end{equation}

Here, $b_n^{0/+}$ are the coefficients of the expansion which are free parameters and they 
obey the unitarity constraint
\begin{equation}
\Sigma(b^{0/+},N_z) \equiv\sum_{m,n=0}^{N_z} B_{mn} ~ b_m^{0/+} b_n^{0/+} \le 1\,,
\end{equation}
where the element $B_{mn}$ satisfies $B_{mn}$ = $B_{nm}$ = $B_{0|m-n|}$, the details for which can be seen from \cite{Lattice:2015tia},\cite{BCL}. The conformal map from $q^2$ to z is given by :
\begin{equation}
z(q^2) = \frac{\sqrt{t_+-q^2}-\sqrt{t_+-t_0}}{\sqrt{t_+-q^2}+\sqrt{t_+-t_0}}\,,  
\end{equation}
where
$t_\pm \equiv (m_B\pm m_{\pi})^2$ and $t_0\equiv t_+(1-\sqrt{1-t_-/t_+})$. $t_0$ is a free parameter that governs the size of $z$ in the semileptonic phase space. For BSZ, the parametrization of any form-factor reads: 
 \begin{equation}\label{eq:bszexp}
 f_i(q^2) = \frac{1}{1 - q^2/m_{R,i}^2} \sum_{k=0}^N a_k^i \, [z(q^2)-z(0)]^k\,,
 \end{equation}
where $m_{R,i}$ denotes the mass of sub-threshold resonances compatible with the quantum numbers of the respective form factors and $a_k^i$s are the coefficients of expansion. The details are provided in \cite{Straub:2015ica}.

The simplified series expansion discussed above is the model-independent description of the form factor and based on analyticity arguments. The z-expansion is a conformal mapping that maps the kinematically allowed region within a disc of radius $|z| < 1$. Only a few coefficients are needed to represent the form factor accurately. Furthermore, it provides a prescription for introducing more parameters with the improvement of data. The BCL expansion obeys the known asymptotic behaviour near the $B\pi$ threshold: $Im(f_+(q^2)) \sim (q^2-t_+)^{3/2}$. Therefore, at $q^2=t_+$ ($z=-1$), the derivative of the form factor must satisfy 
\begin{equation}
\frac{df_+}{dz}\big|_{z=-1} = 0. 
\end{equation}  
BCL uses this constraint to remove an independent degree of freedom from the series expansion in $z$ and arrives at a formula for $f_+$ given in eq.~\ref{eq:bclexpfp}. For the scalar form factor $f_0$ or it's derivative there are no such constraints available at any value of $z$, so we can not remove a further degree of freedom in the series expansion of $f_0(z)$. In contrast, the BSZ form factor parametrization does not obey the known asymptotic behaviour near the $B\pi$ threshold. However, it has the advantage that the value of the form factor at $q^2 = 0$ is among the fit parameters which can be seen from eq.~\ref{eq:bszexp}: $f_i(q^2=0)= a_0^i$. Also, in BSZ the kinematical constraints $f_+(q^2=0) =f_0(q^2=0)$ directly leads to the relation $a_0^+ = a_0^0$ between the coefficients. As is evident from eqs.~\ref{eq:bclexpfp} and \ref{eq:bclexpf0}, in the BCL parametrization the same kinematic constraint leads to a complex relationship between the expansion coefficients: 
\begin{equation}
b_3^0 =  45.70 ( b_0^+ - b_0^0) - 12.78 b_1^0 - 3.58 b_2^0 + 12.85 b_1^+ + 3.44 b_2^+ + 1.21 b_3^+.
\end{equation}
Following this equation, we have replaced $b_3^0$ in terms of the other coefficients in the fit. This helps us reduce one parameter from the fit. 

There is an important difference between the BCL and BSZ parametrization of $f_0(q^2)$ which is due to the presence of the pole factor $\frac{1}{1 - q^2/m_{B^*}^2}$ in BSZ parametrization (eq. \ref{eq:bszexp}) that is absent in the respective expression in BCL (eq. \ref{eq:bclexpf0}). Note that the LCSR pseudo data points are obtained following the parametrization given in eq. \ref{eq:bszexp} with the scalar resonance $M_{B^*} = 5.54$ {\it GeV} \cite{Gubernari:2018wyi} which is slightly above the $B\pi$ threshold. In ref. \cite{BCL}, the authors did not discuss about the parametrization for $f_0(q^2)$. In refs. \cite{Flynn:2015mha,Gelzer:2019zwx}, the functional form of $f_0$ does not include the pole following the argument that the scalar $B^*$ meson whose mass is expected to be above the $B\pi$ threshold has not been observed experimentally. In order to be consistent with the literature, we have followed a functional form for $f_0$ similar to that given in eq. \ref{eq:bclexpf0}. In the result section, we will discuss the impact of this difference on the outcome of our analysis.

\subsubsection{Comparison with existing literature}

\begin{table}[t]
	\centering
	\small
	\begin{ruledtabular}
		\renewcommand*{\arraystretch}{1.3}	
		\begin{tabular}{*{4}{c}}
			$\Delta q^2\left[GeV^2\right]$  &  \multicolumn{3}{c}{$\left(\Delta\mathcal{B}\left(B^0\to\pi^-\ell^+\nu_{\ell}\right)/\Delta q^2\right)~ [10^{-7}]$}  \\
			&  HFLAV Average  &  Our Average & Average (Dropping  BaBar(11)~\cite{delAmoSanchez:2010af}) \\
			\hline
			$\text{0-2}$  &  $\text{72.0$\pm $7.0}$  &  $\text{67.0$\pm $6.4}$ & $64.9\pm 6.7$  \\
			$\text{2-4}$  &  $\text{71.4$\pm $4.6}$  &  $\text{67.4$\pm $4.7}$ & $66.6\pm 4.8$  \\
			$\text{4-6}$  &  $\text{67.0$\pm $3.9}$  &  $\text{65.9$\pm $3.9}$ & $63.7\pm 4.2$ \\
			$\text{6-8}$  &  $\text{75.6$\pm $4.3}$  &  $\text{73.7$\pm $4.2}$  & $69.6\pm 4.6$ \\
			$\text{8-10}$  &  $\text{64.4$\pm $4.3}$  &  $\text{60.2$\pm $4.4}$ &  $64.3\pm 4.7$ \\
			$\text{10-12}$  &  $\text{71.7$\pm $4.6}$  &  $\text{70.4$\pm $4.8}$ &  $73.4\pm 5.2$ \\
			$\text{12-14}$  &  $\text{66.7$\pm $4.7}$  &  $\text{63.0$\pm $4.8}$ & $64.8\pm 5.1$ \\
			$\text{14-16}$  &  $\text{63.3$\pm $4.8}$  &  $\text{61.0$\pm $4.8}$ &  $63.2\pm 5.1$ \\
			$\text{16-18}$  &  $\text{62.0$\pm $4.4}$  &  $\text{60.5$\pm $4.5}$ &  $60.5\pm 4.8$ \\
			$\text{18-20}$  &  $\text{43.2$\pm $4.3}$  &  $\text{41.5$\pm $4.2}$ &  $40.6\pm 4.4$  \\
			$\text{20-22}$  &  $\text{42.5$\pm $4.1}$  &  $\text{39.7$\pm $4.1}$ & $43.2\pm 4.3$ \\
			$\text{22-24}$  &  $\text{34.0$\pm $4.2}$  &  $\text{29.9$\pm $4.4}$ & $33.5\pm 4.7$ \\
			$\text{24-26.4}$  &  $\text{11.7$\pm $2.6}$  &  $\text{10.4$\pm $2.7}$ & $11.8\pm 2.7$   \\
			\hline
			$p$-value &  6\%   &  $\approx$ 1\% &  24.8\% \\
		\end{tabular}
	\end{ruledtabular}
	\caption{\small Comparison between our and HFLAV's average $q^2$ spectrum.} 
	\label{tab:hflav1comp}
\end{table}

In the introduction, we have mentioned the exclusive determination of $|V_{ub}|$ from the available data on $\btopilnu$ decay rates \cite{delAmoSanchez:2010af,Ha:2010rf,
Lees:2012vv,Sibidanov:2013rkk}. The present way of combining all these measurements, followed by HFLAV, is the following:
\begin{itemize}
	\item First, a binned maximum-likelihood fit to determine the average partial branching fraction in each $q^2$ interval is performed. Though the bin-widths used by different measurements are different, the bin-edges are the same (a small difference is mentioned in the next paragraph)\footnote{The highest edge of the highest bin in the most recently published Belle Hadronic-tagged result \cite{Sibidanov:2013rkk} is quoted to be $26~GeV^2$, whereas for the rest of the measurements, it is $26.4~GeV^2$. HFLAV \cite{Amhis:2016xyh,Amhis:2019ckw} seems to be using a different highest-bin-result for $\Delta\mathcal{B}$ than that published by Belle. We, in this analysis, have consistently used the published results from ref. \cite{Sibidanov:2013rkk}.}.
	
	The total branching fraction is then calculated from the sum of the partial branching fractions in the average $q^2$ spectrum, taking the correlations between $q^2$ bins into account. Also, they have treated the systematic and statistical uncertainties and the corresponding correlations separately. In addition, the shared
	sources of systematic uncertainty of all measurements are included in the likelihood as nuisance parameters, for details see \cite{Amhis:2019ckw}. 
	\item In the second stage, the average $q^2$ spectrum is then used to fit $V_{ub}$ and the relevant form-factor parameters (BCL~\cite{Bourrely:2008za} $3+1$: 3 form factor coefficients $b_0$, $b_1$, $b_2$ and $V_{ub}$.). To constrain the high $q^2$ behavior of the spectrum, they use the FLAG lattice average~\cite{Aoki:2016frl} of two LQCD calculations~\cite{Lattice:2015tia,Flynn:2015mha} and similarly, ref.~\cite{Bharucha:2012wy} as LCSR inputs for constraining the low $q^2$ nature of the spectrum.
\end{itemize}

We have repeated the binned maximum-likelihood fit to obtain the average $q^2$-spectrum. Our results, with all the available data, are shown in the third column of table \ref{tab:hflav1comp}. The corresponding correlation matrix of the average $q^2$-spectrum can be found in table~\ref{tab:q2avgcorr} in appendix~\ref{sec:extra_tables}. The results from HFLAV have been displayed in the second column of the same table.  Note that the average spectrum we reproduce is consistent with that from HFLAV within $1 \sigma$. However, our fit quality is about 1\% while that for HFLAV is about 6\%. This difference in the quality of fit could be due to the non-availability of all the information on the shared systematic uncertainties between measurements (like continuum subtraction, tracking efficiency, etc.) in our fit, which HFLAV had utilised in their analysis. We have incorporated the systematic and statistical uncertainties as given in the published papers by Belle and BaBar. 

\begin{figure}[t]
	\centering
	\includegraphics[width=0.55\textwidth]{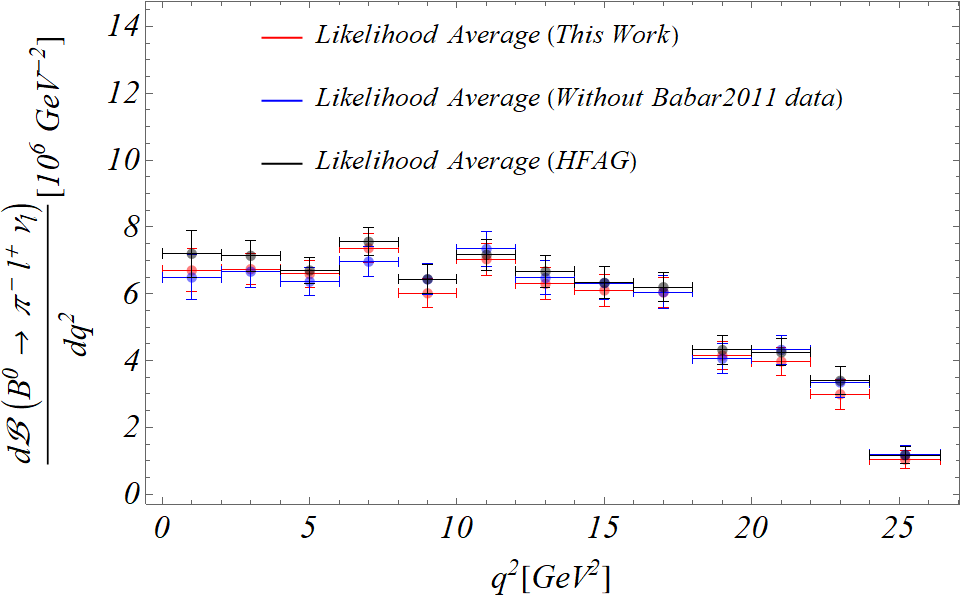}
	\caption{The comparison of our average $q^2$ spectrum of the partial branching fractions in $B\to\pi\ell\nu$ with the one obtained by HFLAV.}
	\label{fig:avgq2comp}
\end{figure}

\begin{table}[t]
	\centering
	\small
	\begin{ruledtabular}
		\renewcommand*{\arraystretch}{1.3}	
		\begin{tabular}{*{5}{c}}
			$\text{Params}$  &  HFLAV Result  &  Our Result & Our Result & From our avg  \\
			&  ($p$-value $= 47\%$)  &  Using HFLAV's avg  &   Using our avg  & Without BaBar(11)~\cite{delAmoSanchez:2010af} \\
			&  &  ($p$-value $= 48\%$)  &   ($p$-value $= 80\%$) &  ($p$-value $= 64\%$) \\
			\hline
			$V_{ub}\times 10^3$  &  $3.67(15)$  &  $3.66(14)$  &  $3.60(12)$ &  $\text{3.66(13)}$  \\
			%	\hline
			$\text{$b^+_0$}$  &  $\text{0.418(12)}$  &  $\text{0.417(12)}$  &  $\text{0.419(12)}$   &  $\text{0.420(12)}$  \\
			%	\hline
			$\text{$b^+_1$}$  &  $\text{-0.399(33)}$  &  $\text{-0.406(31)}$  &  $\text{-0.395(31)}$   &  $\text{-0.410(31)}$  \\
			%	\hline
			$\text{$b^+_2$}$  &  $\text{-0.578(130)}$  &  $\text{-0.605(148)}$  &  $\text{-0.638(150)}$  &  $\text{-0.740(154)}$  \\
		\end{tabular}
	\end{ruledtabular}
	\caption{\small Comparison between our and HFLAV's fit to the average $q^2$ spectrum along with the inputs from lattice and LCSR ($q^2 =0$), see the text for details.} 
	\label{tab:hflav2comp}
\end{table}

In search of a possibility of improvement, one should carefully inspect all the data-sets, which we do in the following sections. A closer look at the data shows that BaBar(12) untagged analysis \cite{Lees:2012vv} of the $B^{0,+}$ modes have much better statistics/yield (almost double) than the one published in the previous year: BaBar(11) \cite{delAmoSanchez:2010af}. It also has the following advantages over the BaBar(11) analysis:
\begin{itemize}
	\item The event selection has been optimized over the entire fit region instead of the signal-enhanced region, as was done previously. 
	\item The tighter selections produce a data-set with a better signal to background ratio and higher purity in the $\btopiplnu$ decays. 
	\item This analysis uses the full BaBar data-set compared to only a subset in the analysis of 2011.
\end{itemize}  

BaBar(11) have presented their main results from a simultaneous analysis of four exclusive charmless semileptonic decay modes: $\btopiplnu$, $\btopi0lnu$, $\btorho0nu$, and $\btorhoplnu$. This method may reduce the cross-feed's sensitivity between $\btopilnu$ and $\btorholnu$ decay modes and some of the background contributions. In 2012, the analysis for $\btopiplnu$ mode was done in 12-bins each of width $\Delta q^2 = 2$ GeV$^2$ (except the last one) and for $\btopi0lnu$ mode in 11-bins. In contrast to this, the study in 2011 was done only in $6$-bins each of width $\Delta q^2 = 4$ GeV$^2$ (except the last one). Note that the analysis-method in BaBar(11) is considerably different from that of BaBar(12). It is also very different from all the analyses on $\btopilnu$ decays by Belle. Thus, as a first attempt to look for the possibility of improvement in the analysis, we drop the BaBar(11) data in $6$-bins while extracting the average partial branching fraction in each $q^2$ interval from a binned maximum-likelihood fit to data. The  result of the fit has been shown in the fourth column in table \ref{tab:hflav1comp}. We notice an improvement in the fit quality from $1\%$ to $24.8\%$. The average $q^2$-spectrum generated in this scenario is consistent with HFLAV and our averages with all data. Figure \ref{fig:avgq2comp} compares all the three average $q^2$-spectrums as given in table \ref{tab:hflav1comp}.  

In the second stage, for the BCL fit of the average $q^2$ spectrum, we are able to more or less reproduce their result for $|V_{ub}|$ and the form-factor parameters (second and third column of table \ref{tab:hflav2comp}), using tables 81 and 82 of ref. \cite{Amhis:2019ckw} \footnote{Here too, there is mismatch between the website and the paper ($p$-value is $61\%$ in website but $47\%$ in the paper) though apparently, there are no differences in the inputs for the fit.}. If we use our own averaged $q^2$ spectrum (all data) instead, the fit-probability increases a lot, but the value of $|V_{ub}|$ decreases (i.e. goes away from the inclusive result) even more (fourth column of table \ref{tab:hflav2comp}). However, from a fit to our $q^2$ average spectrum as obtained without BaBar(11), we get a similar value of $|V_{ub}|$ as HFLAV and the quality of fit is also good with a $p$-value of $64\%$ (fifth and last column of table \ref{tab:hflav2comp}). Note that we have also done a fit after adding the BaBar(11) data-set with this $q^2$-averaged spectrum, the fit quality of which is not good (p-value 3\%).      
   
Other than HFLAV, the lattice group MILC~\cite{Lattice:2015tia} also combined these experimental results. Instead of finding an average $q^2$ spectrum, they fitted the experimental data directly, with or without lattice constraints. It was thus imperative for us to cross-check our analysis with ref. \cite{Lattice:2015tia}. We also reproduced table $XV$ of that paper, where just the experimental data from each measurement are fitted with a BCL parametrization truncated at $n=3$. Table \ref{tab:MILC_comp} compares the results of our fit with those obtained by MILC.

\begin{table}[t]
	\centering
	%\scriptsize
	\begin{ruledtabular}
			\renewcommand*{\arraystretch}{1.3}
	\begin{tabular}{ cccccccc }
		Scenario & $\text{Fit}$  & $\chi ^2\text{/dof}$  &  $\text{dof}$  &  $p$-value $(\%)$  &   $b_1/b_0$  &   $b_2/b_0$  & $b_0|V_{ub}|\times 10^3$  \\
		\hline
		This Work &	$\text{All exp}$  &  $1.4$  &  $48$  &  $3$  &  $\text{-1.05(19)}$  &  $\text{-1.08(60)}$  &  $\text{ 1.52(3)}$  \\
		&$\text{BaBar(11)}$\cite{delAmoSanchez:2010af}  &  $2$  &  $3$  &  $10$  &  $\text{-0.95(43)}$  &  $\text{0.7(1.4)}$  &  $\text{1.36(7)}$  \\
		&$\text{BaBar(12)}$\cite{Lees:2012vv}  &  $0.4$  &  $9$  &  $91$  &  $\text{-0.32(44)}$  &  $\text{-3.6}(1.3)$  &  $\text{1.50(6)}$  \\
		&$\text{Belle(11)}$\cite{Ha:2010rf}  &  $1.2$  &  $10$  &  $30$  &  $\text{-1.32(27)}$  &  $\text{-0.69(91)}$  &  $\text{1.60(6)}$  \\
		&$\text{Belle(13)}$\cite{Sibidanov:2013rkk}  &  $1.3$  &  $17$  &  $19$  &  $\text{-1.89(50)}$  &  $\text{1.4(1.6)}$  &  $\text{1.56(8)}$  \\
		\hline
		MILC~\cite{Lattice:2015tia} &	$\text{All exp}$  &  $1.5$  &  $48$  &  $2$  &  $\text{-0.93(22)}$  &  $\text{-1.54(65)}$  &  $\text{1.53(4)}$  \\
		&$\text{BaBar(11)}$  &  $2$  &  $3$  &  $12$  &  $\text{-0.89(47)}$  &  $\text{0.5(1.5)}$  &  $\text{1.36(7)}$  \\
		&$\text{BaBar(12)}$  &  $1.2$  &  $9$  &  $31$  &  $\text{-0.48(59)}$  &  $\text{-3.2(1.7)}$  &  $\text{1.54(9)}$  \\
		&$\text{Belle(11)}$  &  $1.1$  &  $10$  &  $36$  &  $\text{-1.21(33)}$  &  $\text{-1.18(95)}$  &  $\text{1.63(7)}$  \\
		&$\text{Belle(13)}$  &  $1.2$  &  $17$  &  $23$  &  $\text{-1.89(50)}$  &  $\text{1.4(1.6)}$  &  $\text{1.56(8)}$ \\
	\end{tabular}
	\end{ruledtabular}
	\caption{\small Comparison between table $XV$ of ref.~\cite{Lattice:2015tia} (MILC) and the results of this work. Details are in the text.} 
	\label{tab:MILC_comp}
\end{table}

It is clear from table~\ref{tab:MILC_comp} that our results are in complete agreement with those obtained by MILC within $1\sigma$ for all the cases, except for BaBar(12) where, with consistent parameter-spaces, our fit-probability appears to be much better than that of MILC. In order to further make sure of the validity of our results, we have compared this particular case with a fit to the BaBar(12) data alone that has been provided in table VI of the corresponding reference~\cite{Lees:2012vv} using a BGL expansion for the form factors, the details of which can be obtained from~\cite{delAmoSanchez:2010af} and the references therein. The comparisons are provided in table~\ref{tab:Babar12_comp}. It shows that we are in good agreement with the fit results from BaBar(12).

\begin{table}[t]
	\centering
	%\scriptsize
	\begin{ruledtabular}
		\renewcommand*{\arraystretch}{1.3}
	\begin{tabular}{cccccc}
		Scenario  &  $a_1/a_0$  &  $a_2/a_0$  &  $\chi^2/\text{dof}$  &  $p$-value $(\%)$  &  $\left|V_{ub}~f_+(0)\right|\times 10^4$  \\
		\hline
		$\text{BaBar(12)}$  &  $-0.93\left(19\right)$  &  $-5.4\left(1.0\right)$  &  $\text{4.07/9}$  &  $90.7$  &  $8.7\left(3\right)$ \\
		This Work  &  $-0.91(27)$  &  $-5.54(149)$  &  $\text{3.93/9}$  &  $91.6$  &  $8.6(4)$
	\end{tabular}
	\end{ruledtabular}
	\caption{\small Comparison between fit-results with only the experimental data (combined) between BaBar(12)\cite{Lees:2012vv} and us. We have reproduced the (combined) fit of Table $VI$ of the same reference.} 
	\label{tab:Babar12_comp}
\end{table}  

\subsubsection{Use of the new LCSR inputs}\label{par:newlcsr}

\begin{table}[t] 
	\centering
	\begin{ruledtabular}
		\renewcommand*{\arraystretch}{1.3}
		\begin{tabular}{c|ccccc}
			$q^2$ (GeV$^2$)  &  $-15$  &  $-10$  &  $-5$  &  $0$  &  $5$  \\
			\hline
			$f_+\left(q^2\right)$  &  $0.107\pm 0.035$  &  $0.131\pm 0.043$  &  $0.164\pm 0.055$  &  $0.212\pm 0.073$  &  $0.283\pm 0.101$  \\
			%	\hline
			$f_0\left(q^2\right)$  &  $0.159\pm 0.051$  &  $0.173\pm 0.056$  &  $0.191\pm 0.063$  &  $-$  &  $0.236\pm 0.085$  \\
			%	\hline
		\end{tabular}
	\end{ruledtabular}
	\caption{\small Updated LCSR inputs at values of $q^2$ other than zero \cite{Gubernari:2018wyi} the corresponding correlations are given in table \ref{tab:lcsrcorr}.}
	\label{tab:lcsrinputs}
\end{table}

\begin{table}[ht]
	\centering
	%\scriptsize
	\begin{ruledtabular}
		\renewcommand*{\arraystretch}{1.3}
		\begin{tabular}{cccc}
			$\text{Parameters}$  &  Our Avg. $q^2$ spec. w/o BaBar(11)  &  Our Avg. $q^2$ spec.  &  Our Avg. $q^2$ spec. w/o BaBar(11)  \\
			&  + New Lattice \& LCSR  &  + New Lattice \& LCSR  &  + New Lattice \& LCSR  \\
			&  + BaBar(11) re-introduced  &   &   \\
			&   ($p$ value $= 0.75\%$) &   ($p$ value $= 20.9\%$) & ($p$ value $= 31\%$) \\  
			\hline
			$V_{ub}\times 10^3$  &  $3.78(13)$  	&	$3.78(13)$  	&  $3.89(14)$  \\
			\hline
			$b^+_0$  		&  $0.410(12)$  			&  	$0.410(12)$  	&  $0.408(12)$  \\
			$b^+_1$  		&  $-0.526(44)$  		&  	$-0.526(44)$  	&  $-0.561(46)$  \\
			$b^+_2$  		&  $-0.39(13)$  		&  $-0.39(13)$  	&  $-0.40(13)$  \\
			$b^+_3$  		&  $0.59(24)$  			&  	$0.59(24)$ 	&  $0.59(25)$  \\
			$b^0_0$  		&  $0.540(16)$  		&  	$0.540(16)$   	&  $0.536(16)$  \\
			$b^0_1$  		&  $-1.617(66)$  		&  	$-1.617(66)$ 	&  $-1.647(66)$  \\
			$b^0_2$  		&  $1.294(146)$  		&  	$1.294(146)$  	&  $1.257(146)$
		\end{tabular}
	\end{ruledtabular}
	\caption{\small The extracted values of $|V_{ub}|$ and the other form-factor parameters using our average $q^2$ spectrum of the partial branching fractions and the new lattice and LCSR inputs. Fourth column uses the avg. $q^2$ spectrum without BaBar(11) data (last column of table \ref{tab:hflav1comp}), while the third uses that from the third column of table \ref{tab:hflav1comp} (with all data). In the second column of this table, we repeat the last fit after recombining BaBar(11) data again.}
	\label{tab:averagelcsr}
\end{table}

As was mentioned in the introduction, updated inputs are available for the form-factors $f_+(q^2)$ and $f_0(q^2)$ for the $B\to\pi$ modes at values of $q^2$ other than zero \cite{Gubernari:2018wyi}. They provide the values and covariance matrix for the form factors $f_+(q^2)$ and for $f_T(q^2)$ at $q^2=-15,-10,-5,0,5$ and $f_0(q^2)$ at $q^2=-15, -10,-5, 5$ GeV$^2$ - which are summarized in table \ref{tab:lcsrinputs}. The corresponding correlations are given in the appendix (table \ref{tab:lcsrcorr}). The value of  $f_0$ at $q^2=0$ is obtained via the QCD relation $f_+(0) = f_0(0)$. We utilize these inputs and repeat our second fit to extract $|V_{ub}|$. This fit was performed using the average $q^2$ spectrum given in the fourth column in table \ref{tab:hflav1comp} (without the inputs from BaBar(11)). In addition to the experimental data and LCSR inputs, we have used the lattice inputs from ref.s \cite{Flynn:2015mha,Lattice:2015tia,Gelzer:2019zwx}. In order to further constrain the high $q^2$ behavior, instead of using the FLAG results, we use both of the 2015 lattice results (from the UKQCD~\cite{Flynn:2015mha} and the MILC~\cite{Lattice:2015tia} collaborations) individually
\footnote{We are aware of the 2019 ``preliminary" lattice results from MILC~\cite{Gelzer:2019zwx} but refrain from using the same as we  could not locate a published version of the ``final" results}.

While UKQCD provides synthetic data points for $f_{+,0}(q^2)$ with full covariance matrices (both systematic and statistical) at $q^2 = 19,~22.6,~25.1$ GeV$^2$, MILC only provides the fit-results for their coefficients, using either just lattice or lattice and Experimental data. We use the results of the `only lattice' fit and generate correlated synthetic data-points at exactly the same $q^2$ values as UKQCD, with an extra point for $f_+$ at $q^2 = 20.5$ GeV$^2$, thus utilizing the full information from the lattice fit\footnote{We note at this stage that `using the full fit information' means keeping the number of synthetic data-points equal to the rank of the covariance matrix of the original fit. Consequently, we observe that using the extra $f_+$ input brings down variance of the higher order form-factor-parameters by an order.}.

We find that both the collaborations are in good agreement. Therefore, we use 22 data points (9 from LCSR, 13 from Lattice (3 for each of $f_{+,0}$ from UKQCD, 4 for $f_+$  and 3 for $f_0$ from MILC ) from lattice and LCSR. 
The next obvious step would be to check how the decay distribution fits to the average $q^2$ spectrum from table \ref{tab:hflav1comp}, if these new lattice and LCSR inputs are used. Using the average $q^2$ spectrum from the third and fourth columns of table \ref{tab:hflav1comp} and the 22 data-points from lattice and LCSR, we perform two fits. The results are shown in the last two columns of table \ref{tab:averagelcsr} and it shows that in both the fits the extracted values of $|V_{ub}|$ have increased from that obtained in table \ref{tab:hflav2comp} (fourth and fifth columns, respectively). Unlike the fit following HFLAV, here the $p$-value clearly increases if we use the avg. $q^2$ spectrum without BaBar(11) data. The only problem here is the abysmal fit quality of the avg. $q^2$ fit (third column of table \ref{tab:hflav1comp}). As argued earlier, using the results of a fit with negligible significance leads to biased result. If, instead, we start from the avg. $q^2$ spectrum of the fourth column of table \ref{tab:hflav1comp} (fit w/o BaBar(11); has a considerable fit-probability of $24.8\%$), and reintroduce the data from BaBar(11) during the second fit of decay-rate distribution, along with the new 22 data-points from lattice and LCSR, we end up with the second column of table \ref{tab:averagelcsr}. We can see that the fits find the same parameter space, but with meaningless significance ($\sim 0.8\%$). This reinforces the fact that the data from BaBar(11) is quite at odds with all other data-sets.

Towards the goal of keeping every significant data, while successfully weeding out outliers, we perform a rigorous outlier analysis on the available data-set and discuss how that can improve these observations further.

\begin{table}[t]
	\centering
	%\tiny
	\begin{ruledtabular}
		\renewcommand*{\arraystretch}{1.3}
		\begin{tabular}{cccc|cccc}
			\multicolumn{4}{c|}{BSZ} & \multicolumn{4}{c}{BCL}  \\
			$\chi _{\min }^2\text{/DOF}$  &  $p$-value($\%$) & Parameters & Values & $\chi _{\min }^2\text{/DOF}$  &  $p$-value($\%$) & Parameters & Values \\
			\hline
			$\text{4.48/15}$  &  $99.6$  &  $a_0^{+}$  &  $\text{0.213(22)}$  &  $\text{12.88/15}$  &  $61$  &  $b_0^{+}$  &  $\text{0.396(13)}$  \\
			%	\hline
			&  &  $a_1^{+}$  &  $\text{-0.65(14)}$  & & &  $b_1^{+}$  &  $\text{-0.707(70)}$  \\
			%	\hline
			&  &  $a_2^{+}$  &  $\text{0.263(425)}$  &  & &  $b_2^{+}$  &  $\text{-0.36(18)}$  \\
			%	\hline
			&  &  $a_3^{+}$  &  $\text{0.67(31)}$  &  & &  $b_3^{+}$  &  $\text{0.77(32)}$  \\
			%	\hline
			&  &  $a_1^{0}$  &  $\text{0.41(17)}$  &  & &  $b_0^{0}$  &  $\text{0.521(17)}$  \\
			%	\hline
			&  &  $a_2^{0}$  &  $\text{1.46(51)}$  &  & &  $b_1^{0}$  &  $\text{-1.756(78)}$  \\
			%	\hline
			&  &  $a_3^{0}$  &  $\text{1.78(49)}$  &  & &  $b_2^{0}$  &  $\text{1.15(16)}$  \\
		\end{tabular}
	\end{ruledtabular}
	\caption{Fit results of form factor parameters with only LCSR and lattice input. Used to create the plots in figure \ref{fig:motiv}.} 
	\label{tab:LCSRlatfitres}
\end{table}
\begin{figure}[ht]
	\centering
	\subfloat[]{\includegraphics[width=0.55\textwidth]{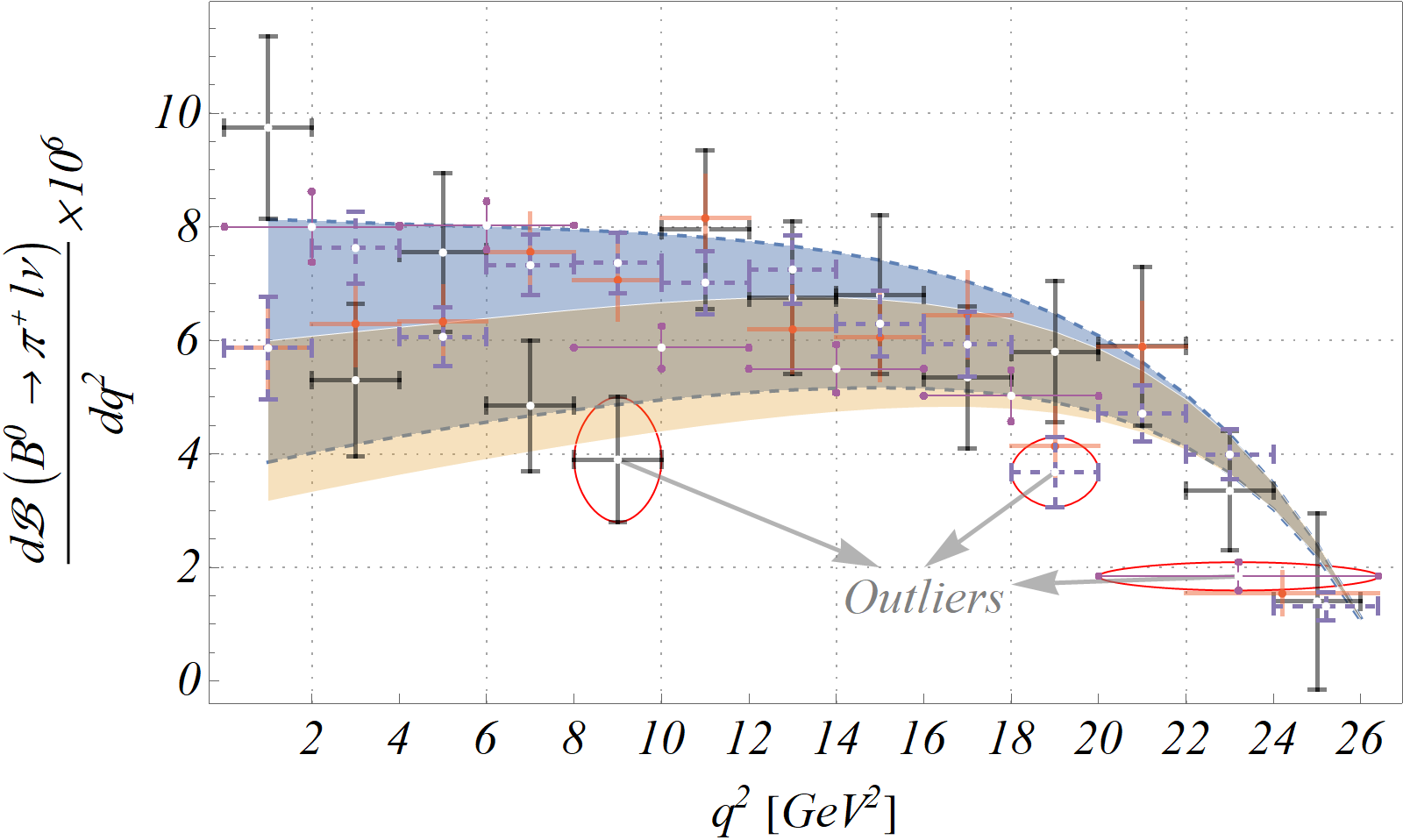}\label{fig:motivB0}}~
	\subfloat[]{\includegraphics[width=0.15\textwidth]{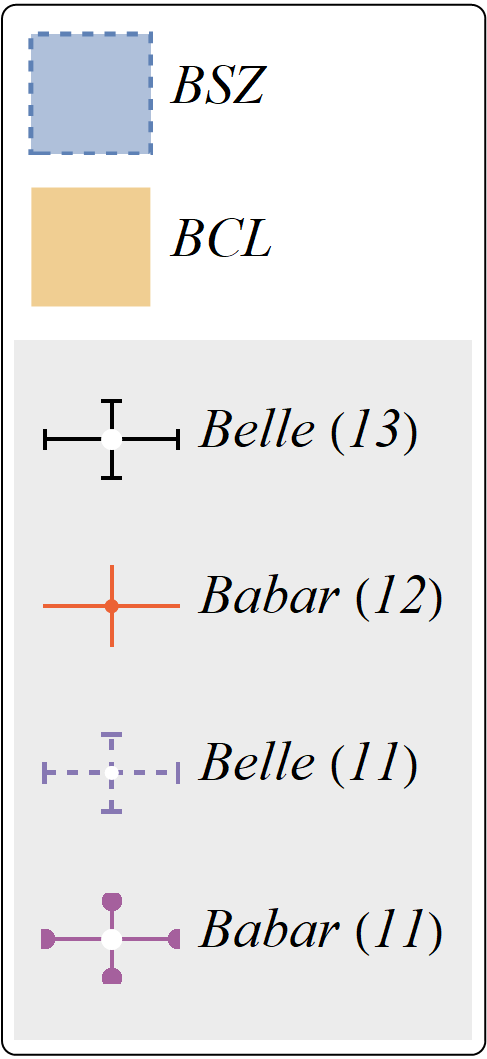}\label{fig:legCombo}}\\
	\subfloat[]{\includegraphics[width=0.55\textwidth]{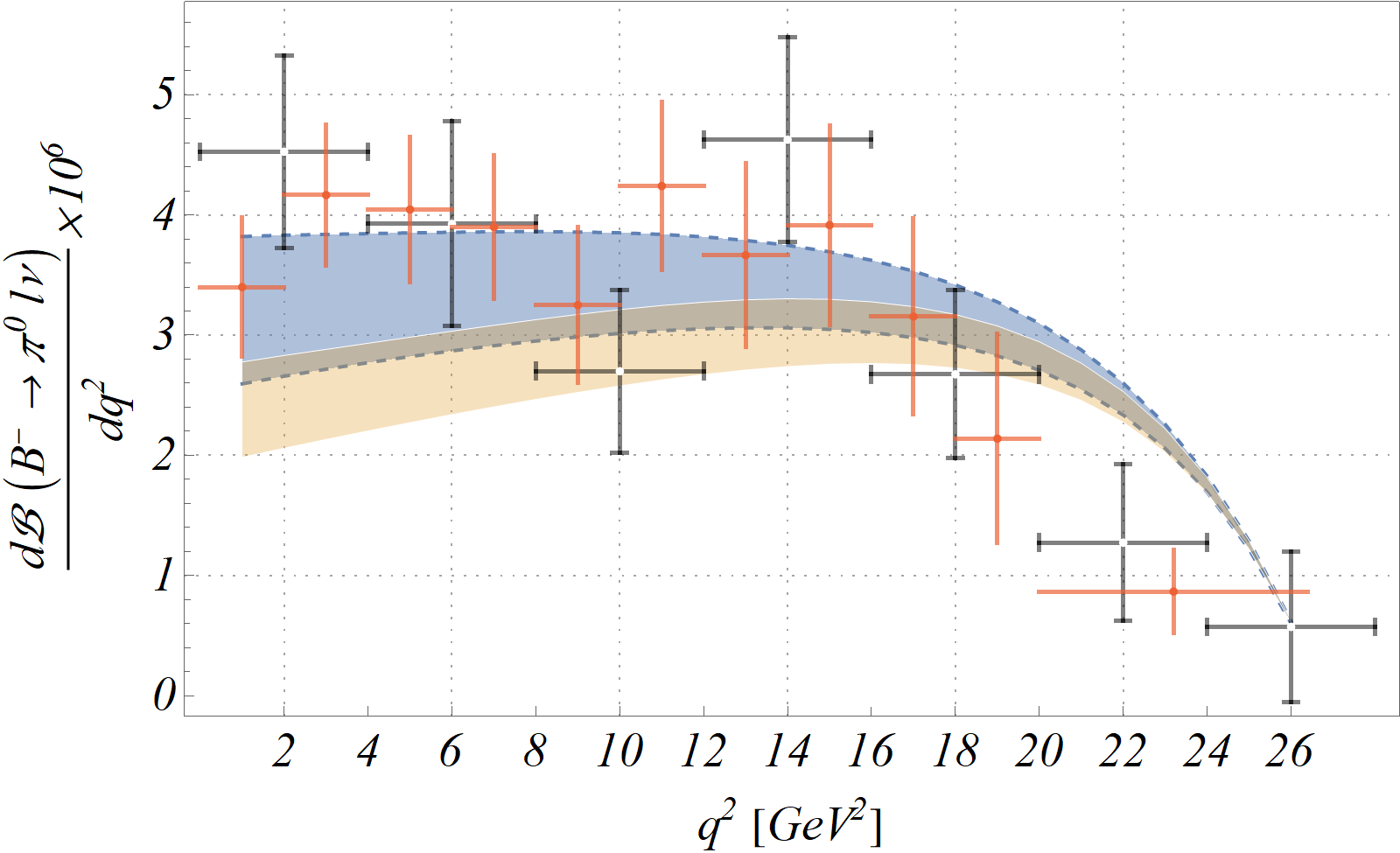}\label{fig:motivBM}}
	\caption{Differential branching fraction plots superposed on experimental data-points, with form factors fitted from lattice and LCSR, and $V_{ub}$ is that obtained from the latest Belle Inclusive Measurement.}
	\label{fig:motiv}
\end{figure}

\subsubsection{$\btopilnu$ rates with $|V_{ub}|^{inc.}$}

In case we neglect the possibility of any new physics effects in $b\to u\ell\nu_{\ell}$ decays then it is natural to expect that the extracted value of $|V_{ub}|$ from the inclusive and exclusive decays should be consistent with each other. At the moment, the extracted values differ from each other by about $2.2~\sigma$. Possible new physics explanation of this observation is available in the literature \cite{Crivellin:2014zpa}. However, in this analysis we will not consider the possibility of new physics in this decay. As we have discussed in the introduction, the extraction of $|V_{ub}|^{inc.}$ is not very clean. Also, it is clear from the discussion above that the values extracted from exclusive $\btopilnu$ decays are very sensitive to a group of experimental inputs, and the fit with all of the data has a minuscule probability. However, we have noted an increase in the extracted value of $|V_{ub}|$ after the inclusion of the new LCSR inputs. 

To understand the effect of the inconsistency in data on the decay rate distributions, we have derived the $\btopilnu$ decay rate distributions using the form-factors extracted only from the LCSR and lattice inputs as discussed above. Armed with the 22 data points from lattice and LCSR we carry out two fits. The first one is with the BSZ (Bharucha, Straub, Zwicky)~\cite{Straub:2015ica} expansion for the form factors. The advantage in using such a series is that the kinematic constraint $f_+(0)=f_0(0)$ is manifest in the way the series is constructed and one need not take care of the same as an extra constraint while fitting. Note that we use the synthetic data provided by the authors wherever we can, in order to minimize the bias from parametrization while using the whole fit-information from the authors.

To find a stable and conservative estimate of the uncertainties of the form-factors, we truncate the series at different orders, starting from 0 to 4 for both $f_+$ and $f_0$ and carry out a model selection procedure incorporating both AIC and AIC${_c}$\footnote{The details about these procedures can be found in our previous articles~\cite{Bhattacharya:2019dot,Biswas:2020uaq}.}. We conclude that the optimal description of the synthetic data is obtained when both $f_0$ and $f_+$ are truncated at $N=3$. This amounts to seven parameters (4 for $f_+$ and 3 for $f_0$). We repeat the same fit with the BCL parametrization as well, with the extra unitarity-constraint applied. The extracted values of the coefficients in BSZ and BCL expansion are given in table \ref{tab:LCSRlatfitres}. We find that with the BCL parametrization, the higher-order form factor parameters are more precise especially for $f_0$. 

To get the shape and height of the distribution, we have used the $|V_{ub}|^{inc.}$ value. This will also help us pinpoint the reason for a discrepancy between the inclusive and exclusive determinations. Using the above fit results and $|V_{ub}|$ from different inclusive estimates, if we calculate the theoretical predictions of the various observables, then any large deviation of those predictions from the actual experimental measurements could potentially diagnose the source of the apparent tension between $|V_{ub}|^{\text{inc}}$ and $|V_{ub}|^{\text{exc}}$.

In figure \ref{fig:motiv}, we have compared the theoretical predictions for the binned differential branching fractions against the experimental information taking all the existing data into account.\footnote{Note that in these plots, we  do not incorporate the data obtained from a combined mode analysis by BaBar 2012 \cite{Lees:2012vv} since they are consistent with the corresponding single-mode analysis.}. We observe that the $q^2$ distribution of the differential branching fraction generated in both the form-factor parametrizations can explain almost all the available data. Very few are lying entirely outside of the theoretical C.I. bands.

In the following section, our goal will be to find out which data-points are picked up as outliers after a rigorous statistical analysis of the exclusive data. Whichever appears both here and that analysis, should by default be responsible for the tension between the inclusive and exclusive estimates of $|V_{ub}|$. Hence, after removing those data-points, we will extract $|V_{ub}|$.

%%%%%%%%%%%%%%%%%%%%%%%%%%%%%%%%%%%%%%%%
\section{Our Main results}
%%%%%%%%%%%%%%%%%%%%%%%%%%%%%%%%%%%%%%%%
%\begingroup
%\squeezetable

\begin{table}[t]
	%	\centering
	\small
	\begin{ruledtabular}
		\renewcommand*{\arraystretch}{1.5}
		\begin{tabular}{ccccccccc}
			Form-  &  Fit  &  $\left[B^0\to \pi^-\right]$  &  $\left[B^0\to \pi^-\right]$  &  $\left[B^0\to \pi^-\right]$  &  $\left[B^0\to \pi^-\right]$  &  $\left[B^0\to \pi^-\right]$  &  $\left[B^0\to \pi^+\right]$  &  $\left[B^0\to \pi^+\right]$  \\
			Factors  &  Index  &  $q^2:~4-8$  &  $q^2:~20-26.4$  &  $q^2:~10-12$  &  $q^2:~20-22$  &  $q^2:~18-20$  &  $q^2:~0.0111637-2$  &  $q^2:~8-10$  \\
			$\text{}$  &  $\text{}$  &  BaBar (11)  &  BaBar (11)  &  BaBar (12)  &  BaBar (12)  &  Belle (11)  &  Belle (13)  &  Belle (13)  \\
			\hline
			$\text{BSZ}$  &  Fit 1A  &  $2.46$  &  $-2.30$  &  $2.08$  &  $-$  &  $-$  &  $-$  &  $-2.42$  \\
			%	\hline
			&  Fit 1B  &  $2.52$  &  $-2.42$  &  $2.07$  &  $-$  &  $-$  &  $-$  &  $-2.41$  \\
			%	\hline
			&  Fit 2A  &  $-$  &  $-$  &  $-$  &  $-$  &  $-2.02$  &  $-$  &  $-2.43$  \\
			%	\hline
			&  Fit 2B  &  $-$  &  $-$  &  $-$  &  $-$  &  $-2.07$  &  $-$  &  $-2.42$  \\
			%	\hline
			&  Fit 3A  &  $2.40$  &  $-2.35$  &  $2.00$  &  $2.01$  &  $-$  &  $-$  &  $-2.44$  \\
			%	\hline
			&  Fit 3B  &  $2.45$  &  $-2.46$  &  $-$  &  $-$  &  $-$  &  $-$  &  $-2.43$  \\
			\hline
			$\text{BCL}$  &  Fit 1A  &  $2.45$  &  $-2.30$  &  $2.07$  &  $-$  &  $-$  &  $-$  &  $-2.42$  \\
			%	\hline
			&  Fit 1B  &  $2.59$  &  $-2.56$  &  $2.07$  &  $-$  &  $-$  &  $-$  &  $-2.40$  \\
			%	\hline
			&  Fit 2A  &  $-$  &  $-$  &  $-$  &  $-$  &  $-2.03$  &  $-$  &  $-2.45$  \\
			%	\hline
			&  Fit 2B  &  $-$  &  $-$  &  $-$  &  $-$  &  $-2.18$  &  $2.00$  &  $-2.42$  \\
			%	\hline
			&  Fit 3A  &  $2.36$  &  $-2.36$  &  $-$  &  $2.00$  &  $-$  &  $-$  &  $-2.45$  \\
			%	\hline
			&  Fit 3B  &  $2.48$  &  $-2.61$  &  $-$  &  $-$  &  $-$  &  $-$  &  $-2.44$  \\
		\end{tabular}
	\end{ruledtabular}
	\caption{\small List of Pulls ($>2~\sigma$) for different fits (with experimental data) in this analysis.} 
	\label{tab:fitpulls}
\end{table}

\begin{table}[t]
	\centering
	\small
	\begin{ruledtabular}
		\renewcommand*{\arraystretch}{1.3}
		\begin{tabular}{c|cccc|cccc}
			\multicolumn{9}{c}{BSZ Parametrization}  \\
			\hline
			Run Name  &  \multicolumn{4}{c|}{Full}  &  \multicolumn{4}{c}{Dropped Pull $>$ 2}  \\
			\hline
			&  $\chi _{\min }^2\text{/DOF}$  &  $p$-value($\%$)  &  \multicolumn{2}{c|}{$V_{ub}\times 10^3$}  &  $\chi _{\min }^2\text{/DOF}$  &  $p$-value($\%$)  &  \multicolumn{2}{c}{$V_{ub}\times 10^3$}  \\
			%		\cline{2-9}
			&  &  &  $\text{Frequentist}$  &  $\text{Bayesian}$  &  &  &  $\text{Freq.}$  &  $\text{Bayes}$  \\
			\hline
			Fit 1A  &  $\text{73.4/56}$  &  $5.92$  &  $\text{3.69(14)}$  &  $\text{3.67(14)}$  &  $\text{46.6/52}$  &  $68.68$  &  $\text{3.79(15)}$  &  $3.77\left(_{16}^{15}\right)$  \\
			%		\hline
			Fit 1B  &  $\text{77./65}$  &  $14.57$  &  $\text{3.74(13)}$  &  $3.73\left(_{14}^{13}\right)$  &  $\text{49.3/61}$  &  $85.77$  &  $\text{3.83(14)}$  &  $3.82\left(_{16}^{14}\right)$  \\
			%		\hline
			Fit 2A  &  $\text{59.5/61}$  &  $53.17$  &  $\text{3.81(14)}$  &  $\text{3.79(15)}$  &  $\text{46./59}$  &  $89.26$  &  $\text{3.86(15)}$  &  $3.85\left(_{16}^{15}\right)$  \\
			%		\hline
			Fit 2B  &  $\text{62./70}$  &  $74.23$  &  $\text{3.85(14)}$  &  $3.83\left(_{15}^{13}\right)$  &  $\text{48.3/68}$  &  $96.63$  &  $\text{3.91(14)}$  &  $3.89\left(_{15}^{14}\right)$  \\
			%		\hline
			Fit 3A  &  $\text{82.2/67}$  &  $9.98$  &  $\text{3.70(14)}$  &  $\text{3.69(14)}$  &  $\text{53.3/62}$  &  $77.56$  &  $\text{3.76(14)}$  &  $3.76\left(_{14}^{15}\right)$  \\
			%		\hline
			Fit 3B  &  $\text{85.9/76}$  &  $20.54$  &  $\text{3.75(13)}$  &  $3.74\left(_{14}^{13}\right)$  &  $\text{62./73}$  &  $81.79$  &  $\text{3.84(14)}$  &  $\text{3.83(14)}$  \\
			\hline\hline
			\multicolumn{9}{c}{BCL Parametrization}  \\
			\hline
			Run Name  &  \multicolumn{4}{c|}{Full}  &  \multicolumn{4}{c}{Dropped Pull $>$ 2}  \\
			\hline
			&  $\chi _{\min }^2\text{/DOF}$  &  $p$-value($\%$)  &  \multicolumn{2}{c|}{$V_{ub}\times 10^3$}  &  $\chi _{\min }^2\text{/DOF}$  &  $p$-value($\%$)  &  \multicolumn{2}{c}{$V_{ub}\times 10^3$}  \\
			%		\cline{2-9}
			&  &  &  $\text{Freq.}$  &  $\text{Bayes}$  &  &  &  $\text{Freq.}$  &  $\text{Bayes}$  \\
			\hline
			Fit 1A  &  $\text{73.5/56}$  &  $5.84$  &  $\text{3.69(14)}$  &  $3.67\left(_{15}^{13}\right)$  &  $\text{46.7/52}$  &  $68.34$  &  $\text{3.79(15)}$  &  $\text{3.78(15)}$  \\
			%		\hline
			Fit 1B  &  $\text{92.1/65}$  &  $1.51$  &  $\text{3.79(13)}$  &  $3.78\left(_{13}^{14}\right)$  &  $\text{63.2/61}$  &  $39.84$  &  $\text{3.89(14)}$  &  $3.87\left(_{15}^{14}\right)$  \\
			%		\hline
			Fit 2A  &  $\text{60.1/61}$  &  $50.8$  &  $\text{3.81(14)}$  &  $\text{3.81(15)}$  &  $\text{46.5/59}$  &  $88.19$  &  $\text{3.87(15)}$  &  $3.85\left(_{15}^{14}\right)$  \\
			%		\hline
			Fit 2B  &  $\text{75.9/70}$  &  $29.42$  &  $\text{3.91(14)}$  &  $\text{3.90(15)}$  &  $\text{58.3/67}$  &  $76.64$  &  $\text{3.96(14)}$  &  $3.96\left(_{14}^{16}\right)$  \\
			%		\hline
			Fit 3A  &  $\text{82.7/67}$  &  $9.35$  &  $\text{3.70(14)}$  &  $3.69\left(_{14}^{13}\right)$  &  $\text{57.8./63}$  &  $66.09$  &  $\text{3.77(14)}$  &  $\text{3.76(15)}$  \\
			%		\hline
			Fit 3B  &  $\text{101.4/76}$  &  $2.73$  &  $\text{3.80(13)}$  &  $3.79\left(_{15}^{13}\right)$  &  $\text{76.3/73}$  &  $37.27$  &  $\text{3.90(14)}$  &  $3.89\left(_{15}^{14}\right)$  \\
		\end{tabular}
	\end{ruledtabular}
	\caption{Freq. and Bayesian} 
	\label{tab:fitresults1}
\end{table}
%\endgroup

We think that instead of using the average $q^2$ spectrum, where the fit clearly is a bad one, we should directly use the individual data-points to do a simultaneous fit of $|V_{ub}|$ and the parameters  corresponding to the chosen form-factor parametrization. This not only allows us to find the outliers, but also takes care of the ambiguity in inferring the results from a two stage fit, one of which is bad and the other good, and provides us with a single value for the fit probability to draw our inference from. We follow the same principle of using individual data-points while using the lattice inputs from different collaborations. Also, in this part of the analysis we have used all the available inputs from LCSR and lattice-QCD as discussed in paragraph \ref{par:newlcsr}. 

To proceed further, we have created different data-sets out of the available experimental inputs from Belle and BaBar. The following are the list of of those data-sets: 
\begin{itemize}
	\item \underline{\bf \textit{Fit 1}}: $B^0$ decays from Belle (2011) and Belle (2013); $B^-$ decays from Belle(2013); the combined modes from BaBar (2011) and BaBar (2012). We have subdivided this set into two sets depending on whether or not LCSR inputs are taken into account, like the following:
	\begin{enumerate}
		\item \emph{\underline{Fit 1A}}: Experimental data (Fit 1) + synthetic Lattice data points, 
		\item \emph{\underline{Fit 1B}}: Experimental data (Fit 1) + synthetic Lattice data points + LCSR.
	\end{enumerate} 
	\item \underline{\bf \textit{Fit 2}}: $B^0$ decays from Belle (2011), BaBar (2012), and Belle (2013); $B^-$ decays from BaBar (2012) and Belle(2013). As above we have defined the following sets:
	\begin{itemize}
		\item \emph{\underline{Fit 2A}}: Experimental data (Fit 2) + synthetic Lattice data points, 
		\item \emph{\underline{Fit 2B}}: Experimental data (Fit 2) + synthetic Lattice data points + LCSR.
	\end{itemize} 
	\item \underline{\bf \textit{Fit 3}}: The combined modes from BaBar (2011) along with the \textit{Fit 2} dataset, with and without LCSR as follows:
	\begin{itemize}
		\item \emph{\underline{Fit 3A}}: Experimental data (Fit 3) + synthetic Lattice data points, 
		\item \emph{\underline{Fit 3B}}: Experimental data (Fit 3) + synthetic Lattice data points + LCSR.
	\end{itemize} 
\end{itemize}
  
{\it Fit 1} contains all data-sets other than the inputs from the single-mode analysis of BaBar(12) \cite{Lees:2012vv}, though we have considered the combined mode analysis from the same publication. Though the partial branching fractions from the analysis of both these modes are consistent with each other, we have also defined the scenario {\it Fit 3} to check their differences. In {\it Fit 3} we have included the single-mode-analysis-data of BaBar(12) and have dropped the combined one. 

Lastly, in {\it Fit 2}, we have dropped the data from the analysis of combined modes from BaBar(11) \cite{delAmoSanchez:2010af} and BaBar(12) \cite{Lees:2012vv}. Basically, this data-set does not contain any input from BaBar(11). As mentioned earlier, this will help us understand the impact of BaBar(11) four-mode analysis data.

For all the fit scenarios we have extracted the respective `pulls' between the data and the fitted distributions. For this work, `pull' for the $i^{\it th}$ data-point will be defined as 
\begin{equation}
pull_i = \frac{\mathcal{O}_i^{exp}-\mathcal{O}_i^{fit}}{\sigma_i^{exp}}\,.
\end{equation}
We have also calculated the Cook's Distances for each data-point for all fits\footnote{Check ref. \cite{Bhattacharya:2019der} and references therein for details on the way this analysis is performed.}. None of the data-points in any of the fits have Cook's distances larger than the Cook-cutoff for that particular fit. This clearly demonstrates that because not a single data-point is tagged as an influential one, the observables with the largest `pull's, would, quite safely, be considered as outliers.

In table \ref{tab:fitpulls}, we present the observables for which the pulls are greater than 2. Note that in all the scenarios, $\mathcal{B}(B^0\to \pi^-)^{[8,10]}$ from Belle(13), and in all the scenarios involving BaBar(11), $\mathcal{B}(B^0\to \pi^-)^{[4,8]}$ and $\mathcal{B}(B^0\to \pi^-)^{[20,26.4]}$ from BaBar(11) have pulls greater than 2. All the fits can more or less comfortably accommodate the rest of the data.

In table \ref{tab:fitresults1}, we have shown the extracted values of $|V_{ub}|$ in different fits. To see the impact of data with pulls $>$ 2, we have repeated all the fits after dropping those respective data-points, shown in the right panel of the same table. We have performed the fit following frequentist as well as Bayesian approaches and compared them. Also, we have done all the fits considering both BCL and BSZ parametrization of the form-factors. The corresponding fit results for the parameters are given in table \ref{tab:fullres}. A few observations are in order here: 
\begin{itemize}
	\item Due to a very-nearly multi-Gaussian nature of the posterior, the results obtained in the frequentist maximum likelihood estimate (MLE) and the medians of the marginal posteriors from the Bayesian analysis are almost identical; in a few cases though, we have obtained slightly asymmetric $1~\sigma$ credible intervals(area between the $\sim 0.16 \to 0.5$, and $\sim 0.5 \to 0.84$ Quantiles of the marginal posterior) in the Bayesian analysis.
	\item For the BSZ parametrization, the quality of fit improves when one includes LCSR, whereas for the BCL case, the fit worsens with the inclusion of LCSR. For \textit{Fit 2}, the fit-probability is reasonably good. This is due to the absence of the BaBar 2011 data set. However, in all the scenarios, the fit quality increases by a considerable amount after dropping a few data-points with pull $>$ 2.
	\item Whenever both Lattice and LCSR data are included, using the BCL form-factor-parametrization results in a slightly larger $|V_{ub}|$ than that obtained from BSZ, albeit with reduced fit-probability. The difference in the best fit values is about $1\%$, and the results are extremely consistent with each other. We will comment on this observation at a later stage. 
	
	\item In all the fits, the extracted $|V_{ub}|$ increases by $\gsim 1\%$ with the inclusion of the new LCSR inputs.
	
	\item We have presented our results truncating the BSZ and BCL expansions at $N=3$~\ref{eq:bszexp} and $N_z=4$~\ref{eq:bclexpfp} respectively, resulting in 4 parameters for $f_+$ and 3 for $f_0$ (due to the kinematic constraint). To check the consistency of the results obtained, we have analysed the available inputs truncating the series at the next order in both the series expansions, i.e., $N=4$ for BSZ and $N_z=5$ for BCL. We found that the extracted values of $|V_{ub}|$ are very much consistent (within their errors) with the one presented in table \ref{tab:fitresults1}. In both types of expansion, we notice a tiny shift $\sim 0.5\%$ in the best fit values of the extracted $|V_{ub}|$. However, in these fits, the newly added higher-order coefficients of the expansions remain mostly unconstrained, and they have a negligible impact on the precision extraction of $|V_{ub}|$. At the present level of precision, it is hard to constrain the higher-order coefficients.  
	
	\item In the scenarios \textit{Fit 1} and \textit{Fit 3}, the extracted values are almost the same, which is not surprising (as mentioned above) since the corresponding data sets are almost-equivalent as well. 
	\item Irrespective of the fit scenario, the extracted $|V_{ub}|$ increases after dropping the data-points with pull $>$ 2. This indicates that the data with large `pull', i.e. those which have a tension with the other data points (as explained earlier, they indeed are the outliers), have an impact on the extracted values of $|V_{ub}|$ too. In \textit{Fit 1}, the extracted $|V_{ub}|$ increases by $\approx$ 3\% whereas the increase is $\gsim 2\%$ in case of \textit{Fit 3}. In \textit{Fit 2}, the enhancement is less (about $1.5\%$), since we had already dropped the BaBar(11) data-set in this case.
	\item The extracted $|V_{ub}|$ in \textit{Fit 2B} is consistent with the one obtained in table \ref{tab:averagelcsr} (fourth column). Hence, the fit results are consistent with the one obtained from our average $q^2$-spectrum (without BaBar(11)) using a BCL parametrization plus the lattice and new LCSR inputs. 
	\item As shall be seen shortly, the partial rates $\mathcal{B}(B^0\to \pi^-)^{[20,26.4]}$ (BaBar(11)) and $\mathcal{B}(B^0\to \pi^-)^{[18,20]}$ (Belle(11)) are extremely sensitive to the extracted value of $|V_{ub}|$. 
\end{itemize}

\begin{figure}[t]
	\centering
	\includegraphics[width=0.8\textwidth]{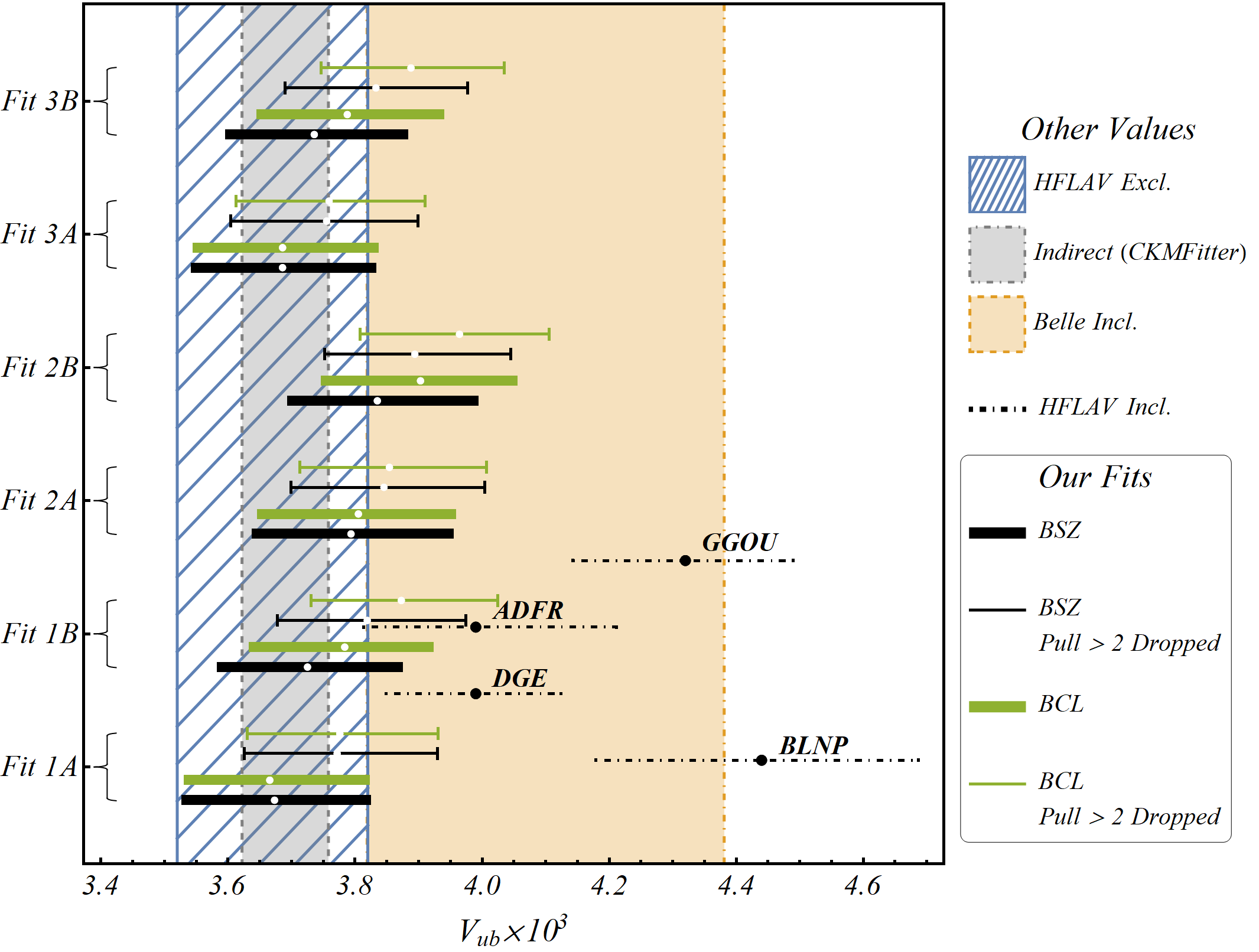}
	\caption{\small Comparison of various $V_{ub}$ results in this work with other measurements.}
	\label{fig:vubcomplot}
\end{figure}

In figure \ref{fig:vubcomplot}, we have compared our extracted values in different fit scenarios as given in table \ref{tab:fitresults1} with the inclusive determinations. Note that our determination for $|V_{ub}|^{exc.}$ is still not consistent at $1~\sigma$ with the values extracted by HFLAV following \textit{GGOU} and \textit{BLNP}. However, they are consistent with the new Belle measurement of $|V_{ub}|^{inc.}$ \cite{Belle:2021xqf}.    
   
\begin{table}[t]
	%	\centering
	\small
	\begin{ruledtabular}
		\renewcommand*{\arraystretch}{1.5}
		\begin{tabular}{cccccccc}
			Form-  &  Inclusive  &  $\left[B^0\to \pi^-\right]$  &  $\left[B^0\to \pi^-\right]$  &  $\left[B^0\to \pi^-\right]$  &  $\left[B^+\to \pi^0\right]$  &  $\left[B^0\to \pi^+\right]$  &  $\left[B^0\to \pi^+\right]$  \\
			Factors  &  $V_{ub}$  &  $q^2:~18-20$  &  $q^2:~20-26.4$  &  $q^2:~18-20$  &  $q^2:~20-26.4$  &  $q^2:~0.0111637-2$  &  $q^2:~8-10$  \\
			$\text{}$  &  used  &  Belle (11)  &  BaBar (11)  &  BaBar (12)  &  BaBar (12)  &  Belle (13)  &  Belle (13)  \\
			\hline
			$\text{BSZ}$  &  HFLAV (GGOU)  &  $-2.55$  &  $-3.54$  &  $-2.13$  &  $-2.35$  &  $-$  &  $-2.04$  \\
			%		\hline
			&  HFLAV (BLNP)  &  $-2.50$  &  $-3.27$  &  $-2.14$  &  $-2.41$  &  $-$  &  $-2.09$  \\
			%		\hline
			&  Belle (New)  &  $-$  &  $-2.32$  &  $-$  &  $-$  &  $2.22$  &  $-$  \\
			\hline
			$\text{BCL}$  &  HFLAV (GGOU)  &  $-2.32$  &  $-3.49$  &  $-$  &  $-2.28$  &  $2.30$  &  $-$  \\
			%		\hline
			&  HFLAV (BLNP)  &  $-2.32$  &  $-3.23$  &  $-$  &  $-2.35$  &  $2.07$  &  $-$  \\
			%		\hline
			&  Belle (New)  &  $-$  &  $-2.29$  &  $-$  &  $-$  &  $2.54$  &  $-$  \\
			\hline
		\end{tabular}
	\end{ruledtabular}
	\caption{List of deviations of theoretical predictions from actual experimental data ($>2~\sigma$).} 
	\label{tab:deviationchart}
\end{table}	   
   
Following up the discussions in section \ref{par:newlcsr}, we perform an analysis where we check the deviations of the data with the predictions of corresponding differential decay rates obtained only from the lattice and LCSR inputs. We have used three different values of $|V_{ub}|^{inc.}$: 
\begin{align}
|V_{ub}| &= (4.10 \pm 0.09\pm 0.22 \pm 0.15)\times 10^{-3},\ \  &\text{ Belle (New) \cite{Belle:2021xqf}}, \nonumber \\ 
|V_{ub}| &= (4.32^{+0.17}_{-0.18})\times 10^{-3} \ &\text{HFLAV (\textit{GGOU}) \cite{Amhis:2019ckw}}, \nonumber\\
|V_{ub}| &= (4.44^{+0.25}_{-0.26})\times 10^{-3}\ \  &\text{HFLAV (\textit{BLNP})} \text{\cite{Amhis:2019ckw}}.
\end{align} 
The deviation corresponding to the $i^{th}$ observable is defined as
\begin{equation}
dev_i = \frac{\mathcal{O}_i^{exp}-\mathcal{O}_i^{SM}}{\sqrt{(\sigma_i^{exp})^2+(\sigma_i^{SM})^2}},
\end{equation}
where $\sigma_i^{SM}$ contains the uncertainties in the decay rates due to form-factor parameters and $|V_{ub}|$. The results of this `deviation'-analysis are shown in table \ref{tab:deviationchart}. As expected, when we use the values of $|V_{ub}|^{inc.}$ from HFLAV, at least four to five data-points have a deviation $>2$. On the other hand, since the new Belle-measurement has a lower value than that of HFLAV, it is only the partial rates $\mathcal{B}(B^0\to \pi^-)^{[20,26.4]}$ (BaBar(11)) and $\mathcal{B}(B^0\to \pi^+)^{[0.01,2]}$ (Belle(13)) which have deviation $>$ 2 (both in BSZ and BCL). However, $\mathcal{B}(B^0\to \pi^+)^{[0.01,2]}$ (Belle(13)) has a rather minor effect on $|V_{ub}|$. The partial decay rates $\mathcal{B}(B^0\to \pi^-)^{[20,26.4]}$ (BaBar(11)), $\mathcal{B}(B^0\to \pi^-)^{[18,20]}$ (Belle(11)) and $\mathcal{B}(B^0\to \pi^-)^{[8,10]}$ (Belle(13)) are the common data points which have pull $>$ 2 in both the analyses given in tables \ref{tab:fitpulls} and \ref{tab:deviationchart}, respectively.

Based on these observations, we define a few additional scenarios: 
\begin{itemize}
	\item \textbf{\textit{\underline{Fit 2B-I}}}: Input used in \textit{Fit 2B} without the data on $\mathcal{B}(B^0\to \pi^-)^{[18,20]}$ (Belle 2011).
	\item \textbf{\textit{\underline{Fit 3B-I}}}: Input used in \textit{Fit 3B} without the data on $\mathcal{B}(B^0\to \pi^-)^{[20,26.4]}$ (BaBar 2011). 
	\item \textbf{\textit{\underline{Fit 3B-II}}}: Input used in \textit{Fit 3B} without the data on $\mathcal{B}(B^0\to \pi^-)^{[18,20]}$ (Belle 2011) and $\mathcal{B}(B^0\to \pi^-)^{[20,26.4]}$ (BaBar 2011). 

\end{itemize}      

\begin{table}
	% \centering
	\small
	\begin{ruledtabular}
		\renewcommand*{\arraystretch}{1.5}
		\begin{tabular}{c|cccc|cccc}
			Fit  &  \multicolumn{4}{c|}{BSZ}  &  \multicolumn{4}{c}{BCL} \\
			Scenario  &  $\chi^2/\text{DOF}$  &  $p$-value($\%$)  &  \multicolumn{2}{c|}{$V_{ub}\times 10^3$}  &  $\chi^2/\text{DOF}$  &  $p$-value($\%$)  &  \multicolumn{2}{c}{$V_{ub}\times 10^3$}  \\
			&  &  &  Frequentist  &  Bayesian  &  &  & Frequentist  &  Bayesian  \\
			\hline
%			Fit F2a  &  $\text{54.08/68}$  &  $89.02$  &  $\text{3.86(13)}$  &  $\text{3.86(13)}$  &  $\text{65.34/68}$  &  $56.91$  &  $\text{3.91(13)}$  &  $\text{3.91(13)}$  \\
			% \hline
			 \textit{F2B-I}  &  $\text{55.4/69}$  &  $88.14$  &  $\text{3.90(14)}$  &  $3.89_{-0.15}^{+0.14}$  &  $\text{68.85/69}$  &  $48.25$  &  $\text{3.96(14)}$  &  $3.95_{-0.15}^{+0.14}$  \\
			% \hline
%			Fit F3a  &  $\text{72.12/73}$  &  $50.7$  &  $\text{3.83(13)}$  &  $\text{3.83(13)}$  &  $\text{84.07/73}$  &  $17.68$  &  $\text{3.88(13)}$  &  $\text{3.88(13)}$  \\
			% \hline
				\textit{F3B-I}  &  $\text{78.86/75}$  &  $35.8$  &  $\text{3.83(14)}$  &  $\text{3.83(13)}$  &  $\text{93.6/75}$  &  $7.19$  &  $\text{3.89(14)}$  &  $\text{3.89(14)}$  \\
			 \textit{F3B-II}  &  $\text{72.96/74}$  &  $51.25$  &  $\text{3.88(14)}$  & $3.87_{-0.15}^{+0.14}$  &  $\text{87.2/74}$  &  $13.99$  &  $\text{3.94(14)}$  &  $3.93_{-0.15}^{+0.14}$  \\
		\end{tabular}
	\end{ruledtabular}
	\caption{Final table of comparison for $|V_{ub}|^{exc.}$ obtained in this work.}
	\label{tab:finalVubRes}
\end{table}

The results in the above-mentioned fit scenarios are given in table \ref{tab:finalVubRes}. A comparison between identical cases, like \textit{Fit 2B} and \textit{Fit 3B}, in tables \ref{tab:fitresults1} and \ref{tab:finalVubRes} shows that one can extract exactly similar values of $|V_{ub}|^{exc.}$ even by dropping only one or two data-points as mentioned above. This means that even in the presence of other outliers, i.e. data-points which do not fit comfortably with other data, the most influential data-points in determining the estimate of $|V_{ub}|^{exc.}$ are the partial branching fractions $\mathcal{B}(B^0\to \pi^-)^{[18,20]}$ (Belle(11)) and $\mathcal{B}(B^0\to \pi^-)^{[20,26.4]}$ (BaBar(11)).   

\begin{figure}[ht]
	\centering
	\subfloat[]{\includegraphics[width=0.5\textwidth]{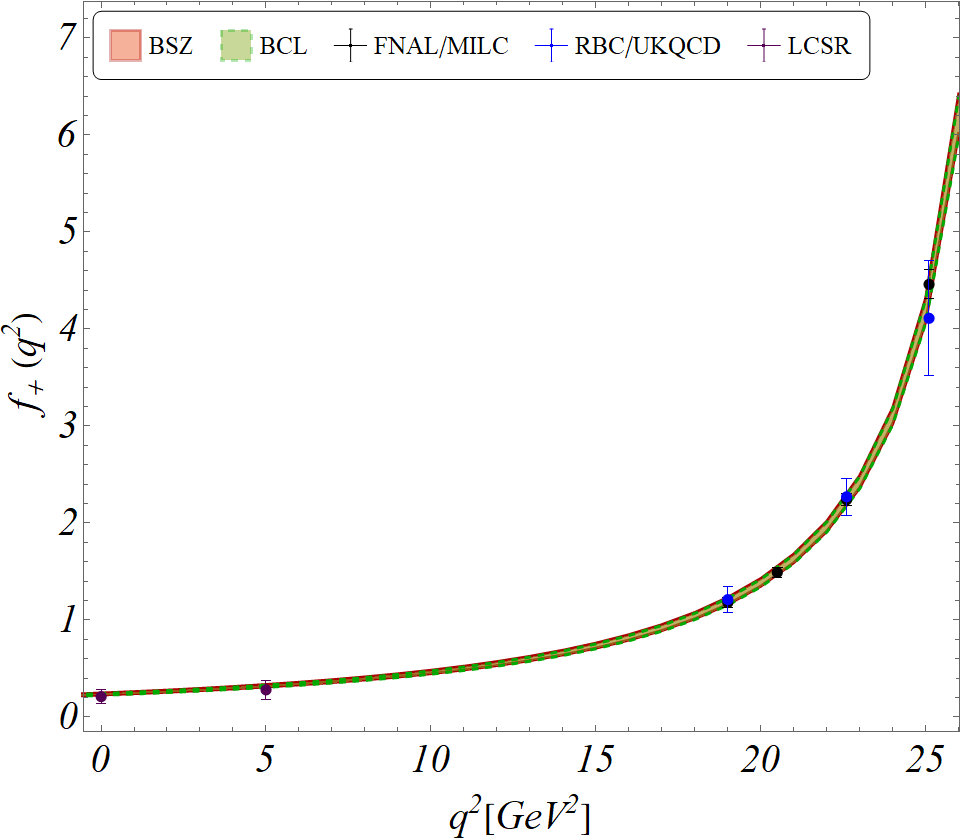}\label{fig:fpluscom}}~~
	\subfloat[]{\includegraphics[width=0.5\textwidth]{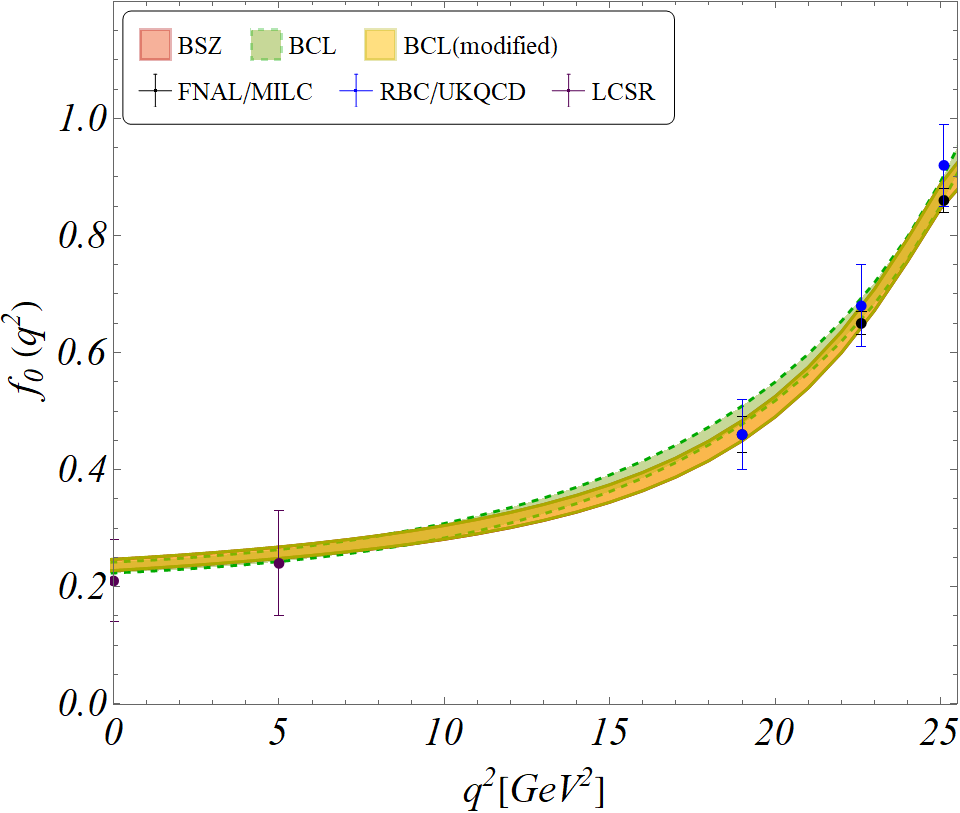}\label{fig:f0com}}
	\caption{Comparison of $q^2$ distributions of the form factors extracted in BSZ and BCL parametrizations. The lattice points and the LCSR pseudo data points at $q^2 = 0$ and $5$ {\it GeV}$^2$ are also shown in the figure.}
	\label{fig:formfactors}
\end{figure}

Finally, we would like to comment on the slight shift ($\approx$ 1\%) in the best fit values of $|V_{ub}|$ in BSZ and BCL expansion of the form factors, though the obtained results are extremely consistent with each other. To understand the difference, we need to compare the form factors obtained from the two separate expansions. As an example, in figure \ref{fig:formfactors} we have compared the form factors calculated from the fit results in `Fit-2B'. Following are a few observations:
\begin{itemize}
	\item The slight difference in the extracted $|V_{ub}|$ is not due to the form factor $f_+(q^2)$, which provides the leading contributions to the decay rate. We can see from figure \ref{fig:fpluscom} that the extracted $q^2$ distributions of $f_+$ in both BSZ and BCL are completely consistent with each other. Also, both distributions satisfy all the lattice or LCSR pseudo data points.
	
	\item The difference in the extracted $|V_{ub}|$ betweent the two formalisms is due to the slight mismatch in the extracted values of $f_0(q^2)$ in a part of the $q^2$ region, which we can see from figure \ref{fig:f0com}. For $15 \lsim q^2 \lsim 23 $ (in {\it GeV$^2$}) the extracted values of $f_0(q^2)$ have a slight mismatch for the two different expansions. However, the extracted values are in good agreement with lattice or LCSR data points in both cases.  
	
	\item  We have checked that the observed difference in $f_0(q^2)$ is because the BSZ expansion uses the pole due to $B^*$ scalar meson while BCL expansion does not. We have discussed it in subsection \ref{subsec:theory}. Our conclusion is based on the following observations: 
	\begin{itemize}
		\item In the fits, after dropping the LCSR data points at $q^2 = 5, -5, -10$ and $-15$ {\it GeV$^2$}, the extracted values of $|V_{ub}|$ in both the expansions exactly match with each other. This indicates the role of the pole factor $1/(1-q^2/M_{B^*}^2)$ in the fit.  
		\item  As a trial, we modified the BCL expansion of $f_0(z)$ in \ref{eq:bclexpf0} after multiplying it by a similar pole factor as given in BSZ expansion and fitted the coefficients using the inputs given in `Fit-2B'. Utilizing this fit result, we have extracted $f_0(q^2)$ and compared it with the BSZ one. The results are shown in figure \ref{fig:f0com} as BCL (modified), and we find absolute agreement between the two results. Also, the extracted values of $|V_{ub}|$ in both the fits become identical. 
		%A similar observation, albeit from a slightly different context, has been made in ref. \cite{Leljak:2021vte}.     
	\end{itemize}     
\end{itemize}   

%%%%%%%%%%%%%%%%%%%%%%%%%%%%%%%%%%%%%%%%
\section{Summary}
%%%%%%%%%%%%%%%%%%%%%%%%%%%%%%%%%%%%%%%%

We have extracted $|V_{ub}|$ analyzing all the available inputs on the exclusive $B\to\pi l\nu$ decays. This includes the data on the partial decay rates, inputs from lattice, and  those from LCSR. In particular, we add the updated inputs on form-factors $f_+(q^2)$ and $f_0(q^2)$ at both zero and non-zero values of $q^2$. We have pointed out some of the issues of the earlier fits done by HFLAV, which relied upon obtaining an average $q^2$ spectrum of the partial width generated from all the available data on the decay rates on $\btopilnu$ in the first stage. To extract $|V_{ub}|$, HFLAV has used this average spectrum at the second stage. We have reproduced both these fits and arrived at a fit with very low probability for the average $q^2$ spectrum at the first stage, similar to HFLAV (ours is even worse). We have identified BaBar(11) data (at least a part of it) as a probable source of such a bad quality fit. The average $q^2$ spectrum of the decay rates without that data-set has an appreciable fit-probability. With this $q^2$ averaged spectrum, in the second stage, we have extracted $|V_{ub}|$ with and without the data from BaBar(11). The quality of fit is much better without the data from BaBar(11). We  have then repeated the same analysis with the new inputs from LCSR and noticed an increase in the best-estimate of $|V_{ub}|$ by roughly about $6\%$. However, the quality of the second-stage-fit is reduced.  

In search of a possibility of improvement, we simultaneously fit all the data (instead of a two-stage fit) after defining different fit scenarios. In the process, we have identified outliers, i.e. data-points inconsistent with the rest of them. The goal is to check whether some of these outliers, if any, are also influential in the extraction of $|V_{ub}|$. We found a very small number of data-points that compromise the fit-quality, and at the same time, influence the extraction of $|V_{ub}|$. Our best results are the following:
\begin{itemize}
	\item Without the input from BaBar(11) (full data-set) and $\mathcal{B}(B^0\to \pi^-)^{[18,20]}$ (Belle(11)), we obtain $|V_{ub}| = (3.95_{-0.15}^{+0.14}) \times 10^{-3}$.
	\item From the full dataset after dropping $\mathcal{B}(B^0\to \pi^-)^{[18,20]}$ (Belle(11)) and $\mathcal{B}(B^0\to \pi^-)^{[20,26.4]}$ (BaBar(11)), the extracted $|V_{ub}| = (3.93_{-0.15}^{+0.14}) \times 10^{-3}$.
\end{itemize}  
Both the values are consistent with the one extracted from inclusive $B\to X_u\ell\nu_{\ell}$ decays within $1~\sigma$.\\

\paragraph*{\textbf{\underline{Note added}}:} During the preparation of this manuscript, a new analysis has come out  which also uses the new LCSR inputs \cite{Leljak:2021vte}, but the average $q^2$ spectrum from HFLAV. They also note a similar trend of obtaining higher values of $|V_{ub}|^{exc.}$ with the new LCSR inputs.  
\vskip 0.5cm

{\bf Acknowledgments:} This work of S.N. is supported by the Science and Engineering Research Board, Govt. of India, under the grant CRG/2018/001260.

\appendix

\section{Extra Tables}\label{sec:extra_tables}

\begin{turnpage}
	\begingroup
	\squeezetable
	\begin{table}
		%	\centering
		\small%\scriptsize
		\begin{ruledtabular}
			\renewcommand*{\arraystretch}{1.5}
			\begin{tabular}{c|cc|cc|cc|cc|cc|cc}
				\multicolumn{13}{c}{BSZ Parametrization}  \\
				\hline
				Params  &  \multicolumn{2}{c|}{Fit 1A}  &  \multicolumn{2}{c|}{Fit 1B}  &  \multicolumn{2}{c|}{Fit 2A}  &  \multicolumn{2}{c|}{Fit 2B}  &  \multicolumn{2}{c|}{Fit 3A} &  \multicolumn{2}{c}{Fit 3B} \\
				\cline{2-13}
				&  $\text{Full}$  &  $\text{Drop.}$  &  $\text{Full}$  &  $\text{Drop.}$  &  $\text{Full}$  &  $\text{Drop.}$  &  $\text{Full}$  &  $\text{Drop.}$  &  $\text{Full}$  &  $\text{Drop.}$  &  $\text{Full}$  &  $\text{Drop.}$  \\
				\hline
				$a_0^{f_+}$  &  $0.254\left(_{11}^{10}\right)$  &  $0.246\left(_{11}^{10}\right)$  &  $0.2493\left(_{99}^{95}\right)$  &  $0.2414\left(_{99}^{95}\right)$  &  $0.241\left(_{11}^{10}\right)$  &  $\text{-0.237(10)}$  &  $0.2374\left(_{98}^{95}\right)$  &  $0.2347\left(_{97}^{94}\right)$  &  $0.254\left(_{11}^{10}\right)$  &  $0.249\left(_{11}^{10}\right)$  &  $0.2496\left(_{98}^{95}\right)$  &  $0.2411\left(_{97}^{94}\right)$  \\
				%		\hline
				$a_1^{f_+}$  &  $\text{-0.61(10)}$  &  $-0.57\left(_{10}^{11}\right)$  &  $-0.614\left(_{89}^{90}\right)$  &  $-0.577\left(_{91}^{92}\right)$  &  $\text{-0.66(11)}$  &  $\text{-0.69(11)}$  &  $-0.649\left(_{92}^{93}\right)$  &  $\text{-0.665(92)}$  &  $-0.601\left(_{100}^{101}\right)$  &  $\text{-0.57(10)}$  &  $-0.606\left(_{88}^{89}\right)$  &  $-0.600\left(_{89}^{90}\right)$  \\
				%		\hline
				$a_2^{f_+}$  &  $-0.0074\left(_{3515}^{3586}\right)$  &  $0.23\left(_{36}^{37}\right)$  &  $0.010\left(_{311}^{315}\right)$  &  $0.23\left(_{32}^{33}\right)$  &  $-0.058\left(_{369}^{375}\right)$  &  $\text{-0.10(37)}$  &  $0.017\left(_{323}^{327}\right)$  &  $-0.011\left(_{321}^{323}\right)$  &  $0.027\left(_{348}^{354}\right)$  &  $0.19\left(_{36}^{37}\right)$  &  $0.037\left(_{310}^{313}\right)$  &  $0.15\left(_{31}^{32}\right)$  \\
				%		\hline
				$a_3^{f_+}$  &  $\text{0.42(26)}$  &  $\text{0.60(27)}$  &  $\text{0.45(23)}$  &  $\text{0.61(24)}$  &  $0.40\left(_{27}^{28}\right)$  &  $\text{0.37(27)}$  &  $\text{0.47(24)}$  &  $\text{0.45(24)}$  &  $\text{0.44(26)}$  &  $\text{0.57(27)}$  &  $\text{0.46(23)}$  &  $0.56\left(_{23}^{24}\right)$  \\
				%		\hline
				$a_1^{f_0}$  &  $\text{0.74(12)}$  &  $\text{0.68(12)}$  &  $\text{0.65(10)}$  &  $0.60\left(_{11}^{10}\right)$  &  $\text{0.64(12)}$  &  $\text{0.61(12)}$  &  $\text{0.56(11)}$  &  $0.54\left(_{11}^{10}\right)$  &  $\text{0.74(12)}$  &  $\text{0.70(12)}$  &  $\text{0.65(10)}$  &  $\text{0.59(10)}$  \\
				%		\hline
				$a_2^{f_0}$  &  $\text{2.34(43)}$  &  $\text{2.20(43)}$  &  $\text{2.01(39)}$  &  $\text{1.91(39)}$  &  $\text{2.08(43)}$  &  $\text{2.01(43)}$  &  $\text{1.82(39)}$  &  $\text{1.77(39)}$  &  $\text{2.35(43)}$  &  $\text{2.26(43)}$  &  $\text{2.01(39)}$  &  $\text{1.89(39)}$  \\
				%		\hline
				$a_3^{f_0}$  &  $\text{2.54(45)}$  &  $\text{2.43(45)}$  &  $\text{2.21(41)}$  &  $\text{2.14(41)}$  &  $\text{2.33(45)}$  &  $\text{2.27(45)}$  &  $\text{2.06(41)}$  &  $\text{2.02(41)}$  &  $\text{2.55(45)}$  &  $\text{2.48(45)}$  &  $\text{2.22(41)}$  &  $\text{2.12(41)}$  \\
				%				\hline
				%				$V_{ub}\times 10^{3}$  &  $3.64(12)$  &  $3.74(12)$  &  $3.68(12)$  &  $3.78(12)$  &  $3.76(12)$  &  $3.82(12)$  &  $3.80(12)$  &  $3.86(12)$  &  $3.65(12)$  &  $3.72(12)$  &  $3.70(12)$  &  $3.79\left(_{13}^{12}\right)$  \\
				\hline
				\multicolumn{13}{c}{BCL Parametrization}  \\
				\hline
				Params  &  \multicolumn{2}{c|}{Fit 1A}  &  \multicolumn{2}{c|}{Fit 1B}  &  \multicolumn{2}{c|}{Fit 2A}  &  \multicolumn{2}{c|}{Fit 2B}  &  \multicolumn{2}{c|}{Fit 3A} &  \multicolumn{2}{c}{Fit 3B} \\
				\cline{2-13}
				&  $\text{Full}$  &  $\text{Drop.}$  &  $\text{Full}$  &  $\text{Drop.}$  &  $\text{Full}$  &  $\text{Drop.}$  &  $\text{Full}$  &  $\text{Drop.}$  &  $\text{Full}$  &  $\text{Drop.}$  &  $\text{Full}$  &  $\text{Drop.}$  \\
				\hline
				$b_0^{f_+}$  &  $\text{0.416(12)}$  &  $\text{0.410(12)}$  &  $\text{0.410(12)}$  &  $\text{0.405(12)}$  &  $\text{0.413(12)}$  &  $\text{0.413(12)}$  &  $\text{0.408(12)}$  &  $\text{0.408(12)}$  &  $\text{0.415(12)}$  &  $\text{0.413(12)}$  &  $\text{0.410(12)}$ & $\text{0.406(12)}$ \\
				%		\hline
				$b_1^{f_+}$  &  $-0.509\left(_{48}^{47}\right)$  &  $-0.559\left(_{50}^{49}\right)$  &  $\text{-0.525(44)}$  &  $\text{-0.575(46)}$  &  $-0.539\left(_{50}^{49}\right)$  &  $\text{-0.542(49)}$  &  $-0.560\left(_{46}^{45}\right)$  &  $-0.562\left(_{46}^{45}\right)$  &  $\text{-0.513(47)}$  &  $-0.538\left(_{50}^{49}\right)$  &  $-0.527\left(_{44}^{43}\right)$  &  $\text{-0.566(45)}$  \\
				%		\hline
				$b_2^{f_+}$  &  $\text{-0.36(14)}$  &  $\text{-0.27(15)}$  &  $\text{-0.39(13)}$  &  $\text{-0.30(13)}$  &  $\text{-0.38(15)}$  &  $\text{-0.41(15)}$  &  $\text{-0.39(13)}$  &  $\text{-0.43(13)}$  &  $\text{-0.34(14)}$  &  $-0.32\left(_{14}^{15}\right)$  &  $\text{-0.38(13)}$  &  $\text{-0.33(13)}$  \\
				%		\hline
				$b_3^{f_+}$  &  $\text{0.44(27)}$  &  $0.63\left(_{27}^{28}\right)$  &  $\text{0.59(24)}$  &  $0.76\left(_{24}^{25}\right)$  &  $\text{0.43(28)}$  &  $\text{0.40(28)}$  &  $\text{0.61(24)}$  &  $0.55\left(_{24}^{25}\right)$  &  $0.47\left(_{26}^{27}\right)$  &  $\text{0.53(27)}$  &  $0.61\left(_{23}^{24}\right)$  &  $\text{0.71(24)}$  \\
				%		\hline
				$b_0^{f_0}$  &  $\text{0.507(18)}$  &  $\text{0.506(18)}$  &  $\text{0.540(16)}$  &  $\text{0.537(16)}$  &  $\text{0.506(18)}$  &  $\text{0.506(18)}$  &  $\text{0.537(16)}$  &  $\text{0.536(16)}$  &  $\text{0.507(18)}$  &  $\text{0.507(18)}$  &  $\text{0.540(16)}$  &  $\text{0.538(16)}$  \\
				%		\hline
				$b_1^{f_0}$  &  $\text{-1.750(73)}$  &  $\text{-1.765(73)}$  &  $-1.616\left(_{66}^{65}\right)$  &  $\text{-1.640(66)}$  &  $\text{-1.768(73)}$  &  $-1.771\left(_{74}^{73}\right)$  &  $\text{-1.643(66)}$  &  $\text{-1.651(66)}$  &  $-1.750\left(_{74}^{73}\right)$  &  $\text{-1.760(73)}$  &  $\text{-1.615(66)}$  &  $\text{-1.638(66)}$  \\
				%		\hline
				$b_2^{f_0}$  &  $\text{1.67(17)}$  &  $\text{1.64(17)}$  &  $\text{1.29(15)}$  &  $\text{1.29(15)}$  &  $\text{1.60(17)}$  &  $\text{1.58(17)}$  &  $\text{1.26(15)}$  &  $\text{1.25(15)}$  &  $\text{1.67(17)}$  &  $\text{1.64(17)}$  &  $1.30\left(_{14}^{15}\right)$  &  $\text{1.28(15)}$  \\
				%				\hline
				%				$V_{ub}\times 10^{3}$  &  $3.64(12)$  &  $3.74(12)$  &  $3.72(12)$  &  $3.82(12)$  &  $3.77(12)$  &  $3.82(12)$  &  $3.85(12)$  &  $3.91(12)$  &  $3.66(12)$  &  $3.72(12)$  &  $3.74(12)$  &  $3.84\left(_{13}^{12}\right)$  \\
			\end{tabular}
		\end{ruledtabular}
		\caption{Full Results of form-factor parameters (Bayesian).} 
		\label{tab:fullres}
	\end{table}
	\endgroup
\end{turnpage}

\begin{table}[ht]
	\centering
	%\tiny
	\begin{ruledtabular}
		\renewcommand*{\arraystretch}{1.3}
		\begin{tabular}{c|ccccccccccccc}
			$\Delta q^2\left[GeV^2\right]$  &  $\text{0-2}$  &  $\text{2-4}$  &  $\text{4-6}$  &  $\text{6-8}$  &  $\text{8-10}$  &  $\text{10-12}$  &  $\text{12-14}$  &  $\text{14-16}$  &  $\text{16-18}$  &  $\text{18-20}$  &  $\text{20-22}$  &  $\text{22-24}$  &  $\text{24-26.4}$  \\
			\hline
			$\text{0-2}$  &  $1.$  &  $-0.24$  &  $0.15$  &  $0.05$  &  $0.14$  &  $0.13$  &  $0.12$  &  $0.09$  &  $0.06$  &  $0.04$  &  $0.07$  &  $0.$  &  $0.01$  \\
			%	\hline
			$\text{2-4}$  &  &  $1.$  &  $-0.14$  &  $0.27$  &  $0.05$  &  $0.11$  &  $0.07$  &  $0.06$  &  $0.04$  &  $0.02$  &  $0.06$  &  $0.04$  &  $0.03$  \\
			%	\hline
			$\text{4-6}$  &  &  &  $1.$  &  $-0.21$  &  $0.22$  &  $0.09$  &  $0.14$  &  $0.08$  &  $0.09$  &  $0.07$  &  $0.09$  &  $0.03$  &  $0.04$  \\
			%	\hline
			$\text{6-8}$  &  &  &  &  $1.$  &  $-0.08$  &  $0.26$  &  $0.13$  &  $0.11$  &  $0.11$  &  $0.05$  &  $0.09$  &  $0.07$  &  $0.04$  \\
			%	\hline
			$\text{8-10}$  &  &  &  &  &  $1.$  &  $-0.23$  &  $0.28$  &  $0.1$  &  $0.05$  &  $0.09$  &  $0.09$  &  $0.01$  &  $0.01$  \\
			%	\hline
			$\text{10-12}$  &  &  &  &  &  &  $1.$  &  $0.02$  &  $0.13$  &  $0.15$  &  $0.02$  &  $0.08$  &  $0.08$  &  $0.04$  \\
			%	\hline
			$\text{12-14}$  &  &  &  &  &  &  &  $1.$  &  $-0.23$  &  $0.15$  &  $0.05$  &  $0.08$  &  $0.02$  &  $0.02$  \\
			%	\hline
			$\text{14-16}$  &  &  &  &  &  &  &  &  $1.$  &  $0.01$  &  $0.16$  &  $0.07$  &  $0.02$  &  $0.02$  \\
			%	\hline
			$\text{16-18}$  &  &  &  &  &  &  &  &  &  $1.$  &  $-0.19$  &  $0.1$  &  $0.08$  &  $0.07$  \\
			%	\hline
			$\text{18-20}$  &  &  &  &  &  &  &  &  &  &  $1.$  &  $0.01$  &  $-0.01$  &  $-0.02$  \\
			%	\hline
			$\text{20-22}$  &  &  &  &  &  &  &  &  &  &  &  $1.$  &  $-0.18$  &  $-0.03$  \\
			%	\hline
			$\text{22-24}$  &  &  &  &  &  &  &  &  &  &  &  &  $1.$  &  $-0.24$  \\
			%	\hline
			$\text{24-26.4}$  &  &  &  &  &  &  &  &  &  &  &  &  &  $1.$  \\
		\end{tabular}
	\end{ruledtabular}
	\caption{\small Correlation matrix of the $q^2$ average spectrum in our analysis. Corresponding to column three of table \ref{tab:hflav1comp}.} 
	\label{tab:q2avgcorr}
\end{table}

\begin{table}[ht]
	\centering
	%\tiny
	\begin{ruledtabular}
		\renewcommand*{\arraystretch}{1.3}
\begin{tabular}{|*{10}{c|}}
%	\hline
	$f(q^2)$ & $f_+\text{(-15)}$ &  $f_+\text{(-10)}$ & $f_+\text{(-5)}$ & $f_+\text{(0)}$ & $f_+\text{(5)}$ & $f_0\text{(-15)}$ & $f_0\text{(-10)}$ & $f_0\text{(-5)}$ & $f_0\text{(5)}$\\
	\hline
	$f_+\text{(-15)}$  &  $1.$  &  $0.77$  &  $0.77$  &  $0.78$  &  $0.78$  &  $0.76$  &  $0.76$  &  $0.77$  &  $0.79$  \\
%	\hline
	$f_+\text{(-10)}$  &  $0.77$  &  $1.$  &  $0.77$  &  $0.78$  &  $0.79$  &  $0.76$  &  $0.77$  &  $0.77$  &  $0.79$  \\
%	\hline
	$f_+\text{(-5)}$  &  $0.77$  &  $0.77$  &  $1.$  &  $0.78$  &  $0.79$  &  $0.77$  &  $0.77$  &  $0.77$  &  $0.79$  \\
%	\hline
	$f_+\text{(0)}$  &  $0.78$  &  $0.78$  &  $0.78$  &  $1.$  &  $0.79$  &  $0.77$  &  $0.77$  &  $0.78$  &  $0.8$  \\
%	\hline
	$f_+\text{(5)}$  &  $0.78$  &  $0.79$  &  $0.79$  &  $0.79$  &  $1.$  &  $0.78$  &  $0.78$  &  $0.79$  &  $0.81$  \\
%	\hline
	$f_0\text{(-15)}$  &  $0.76$  &  $0.76$  &  $0.77$  &  $0.77$  &  $0.78$  &  $1.$  &  $0.76$  &  $0.76$  &  $0.78$  \\
%	\hline
	$f_0\text{(-10)}$  &  $0.76$  &  $0.77$  &  $0.77$  &  $0.77$  &  $0.78$  &  $0.76$  &  $1.$  &  $0.77$  &  $0.78$  \\
%	\hline
	$f_0\text{(-5)}$  &  $0.77$  &  $0.77$  &  $0.77$  &  $0.78$  &  $0.79$  &  $0.76$  &  $0.77$  &  $1.$  &  $0.79$  \\
%	\hline
	$f_0\text{(5)}$  &  $0.79$  &  $0.79$  &  $0.79$  &  $0.8$  &  $0.81$  &  $0.78$  &  $0.78$  &  $0.79$  &  $1.$  \\
%	\hline
\end{tabular}
\end{ruledtabular}
\caption{Correlation matrix of $f_+(q^2)$ and $f_0(q^2)$ at different values of $q^2$ from LCSR \cite{Gubernari:2018wyi}.}
\label{tab:lcsrcorr}
\end{table}

\bibliography{ref_ASSI.bib}
\end{document}